\documentclass[12pt]{article} 
\pdfoutput=1 
\usepackage{jcapmod} 
\usepackage{booktabs}
\usepackage{mathtools}
\usepackage[english]{babel}
\usepackage{amsmath,amssymb,amsbsy,amstext, amsthm}
\usepackage[usenames,dvipsnames]{xcolor}
\usepackage{hyperref}
\hypersetup{
    urlcolor=NavyBlue,
    citecolor=NavyBlue,
    linkcolor=NavyBlue,
}%
\usepackage{array}
\usepackage{url} 
\usepackage{slashed}
\usepackage{multicol}
\usepackage{blkarray}
\usepackage{enumerate}
\usepackage{graphicx}
\usepackage{amsfonts}
\usepackage{amssymb}
  \usepackage{empheq}
  \usepackage{tcolorbox}
\usepackage{xcolor}
\usepackage{soul}
\usepackage{colortbl}

\usepackage{float}
\usepackage[utf8]{inputenc}
\RequirePackage{color}
\usepackage[normalem]{ulem}

\newcommand{\ml}[1]{{\ensuremath{\,\scalebox{0.8}{(#1)}}}}

\usepackage{colortbl}
\definecolor{green2}{cmyk}{0, 1, 0.5, 0}
\definecolor{lightgreen}{cmyk}{0.2, 0, 0.2, 0.2}
\definecolor{lightgray}{cmyk}{0.1,0.2,0,0.1}
\definecolor{lightgray2}{cmyk}{0.4,0.4,0,0.8}
\definecolor{black}{cmyk}{1.0,1.0,1.0,1.0}

\allowdisplaybreaks[1]

\usepackage{colortbl}
\definecolor{lightgreen}{cmyk}{0.2, 0, 0.2, 0.2}
\definecolor{lightgray}{cmyk}{0.1,0.2,0,0.1}
\definecolor{lightgray2}{cmyk}{0.1,0.1,0,0.1}

\makeatletter
\newlength{\apb@width}
\newcommand{\autoparbox}[2][c]{\settowidth{\apb@width}{#2}\parbox[#1]{\apb@width}{#2}}

\makeatother

\numberwithin{equation}{section}

\def\be{\begin{equation}}
\def\ee{\end{equation}}

\def\bea{\begin{eqnarray}}
\def\eea{\end{eqnarray}}

\def\lp{\left(}
\def\rp{\right)}

\def\lb{\left[}
\def\rb{\right]}

\def\del{\partial}
\def\Mp{M_{\rm Pl}}

\def\0{{\boldsymbol 0}}

\usepackage{setspace} 
\begin{document}

\begin{titlepage}

\setcounter{page}{1} \baselineskip=15.5pt \thispagestyle{empty}

\bigskip\

\vspace{1cm}
\begin{center}

{\fontsize{20}{28}\selectfont  \sffamily \bfseries {Cosmological tests of quintessence \\ \vskip0.2cm in  quantum gravity}}

\end{center}

\vspace{-0.1cm}

\begin{center}
{\fontsize{13}{30}\selectfont Sukannya Bhattacharya$^{a,}$\footnote{\texttt{sukannya.bhattacharya@ift.csic.es}}, Giulia Borghetto$^{b,\!}$
\footnote{\texttt{giulia.borghetto@gmail.com}}, Ameek Malhotra$^{b,\!}$ \footnote{\texttt{ameek.malhotra@swansea.ac.uk}}, \\
Susha Parameswaran$^{c,}$\footnote{\texttt{susha.parameswaran@liverpool.ac.uk}},
Gianmassimo Tasinato$^{b,d,}$\footnote{\texttt{g.tasinato2208@gmail.com}},  Ivonne Zavala$^{b,}$\footnote{\texttt{e.i.zavalacarrasco@swansea.ac.uk}}
} 
\end{center}

\begin{center}

\vskip 6pt
\textsl{$^a$ Instituto de F\'isica T\'eorica UAM/CSIC, Calle Nicol\'as Cabrera 13-15, Cantoblanco, 28049, Madrid, Spain}\\
\textsl{$^b$ Physics Department, Swansea University, SA2 8PP, UK}\\
\textsl{$^c$ Department of Mathematical Sciences, University of Liverpool, Liverpool, 
L69 7ZL, UK}\\
\textsl{$^d$ Dipartimento di Fisica e Astronomia, Universit\`a di Bologna and \\
 INFN, Sezione di Bologna, I.S. FLAG, viale B. Pichat 6/2, 40127 Bologna,   Italy
}
\vskip 4pt

\end{center}

\vspace{0.7cm}
\hrule \vspace{0.3cm}
\noindent
We  use a suite of the most  recent cosmological observations to test   models of dynamical dark energy  motivated by  quantum gravity.  Specifically, we focus on hilltop quintessence scenarios,  able to  satisfy theoretical constraints from  quantum gravity. We discuss their  realisation based on axions, their supersymmetric partners, and  Higgs-like string constructions, including dynamical mechanisms to set up initial conditions at the hilltops.  We also examine a specific parameterisation for dynamical dark energy suitable for hilltop quintessence.
   We then perform an analysis based on Markov Chain Monte-Carlo  to assess  
 their predictions against  CMB, galaxy surveys, and supernova data.
 We show  to what extent current data can distinguish amongst different hilltop set-ups, providing model parameter constraints that are complementary to and synergetic with theoretical bounds from quantum gravity conjectures, as well as model comparisons across the main dark energy candidates in the literature. 
However, all these constraints are sensitive to priors based on theoretical assumptions about viable regions of parameter space.  Consequently,  we discuss  theoretical challenges  in refining these priors, with the aim of maximizing the informative 
power of current and forthcoming cosmological datasets for testing  dark energy scenarios in quantum gravity.

\vskip 10pt
\hrule
 
\vspace{0.4cm}

\enlargethispage{\baselineskip}
 \end{titlepage}

 \tableofcontents

\section{Introduction}

One of the most significant and challenging problems in contemporary fundamental physics is
to understand the microscopic nature of the Dark Energy (DE) that dominates our Universe today, driving its current accelerated expansion.  From one perspective, DE may appear  a low-energy problem, since the 
observed DE scale  is small, lying  around the milli-eV.
 However, the fact that vacuum energy -- an ultraviolet sensitive quantum phenomenon --  behaves as DE once included in Einstein's  General Relativity, frames the problem as a high-energy one.
  This makes it all the more exciting that cosmological observations are probing the behaviour of DE with an ever-increasing degree of precision, possibly   opening a path to connect quantum gravity to observations.  In this paper, we consider classes of quintessence models for DE that are currently allowed by quantum gravity considerations,  and we test them against the most recent cosmological data. At the same time, we 
  identify trends that current cosmological results indicate
  for DE model building in quantum gravity.  

String theory provides an excellent framework for the DE problem
 (for  reviews, see e.g.~\cite{Cicoli:2018kdo, Cicoli:2023opf}).  The simplest candidate for DE has long been considered to be a positive vacuum energy (corresponding to a de Sitter vacuum), which, however, must be fine-tuned to a level of one
part in $10^{120}$ -- the so-called cosmological constant problem.  The supposed String Landscape of exponentially large numbers of finely-spaced metastable de Sitter vacua, together with eternal inflation to populate them, lended itself to an anthropic explanation of this fine-tuning (see e.g. \cite{Polchinski:2006gy} for a review).  However, despite impressive technical progress, after two decades of effort there is still no consensus on a single example of an explicit, well-controlled de Sitter vacuum in string theory.  Instead, a number of obstructions are invariably met in the hunt for metastable de Sitter string vacua, including challenges in satisfying global and local constraints, tachyonic instabilities, and a lack of parametric or even numerical control in the perturbative expansions used.  

At the same time, there has been a growing focus on the expectation that, in string theory and quantum gravity, not everything goes: not every effective field theory (EFT) can be ultraviolet completed into a theory of quantum gravity and   those EFTs that are not consistent with quantum gravity are deemed to be in the Swampland.  In mapping out which EFTs lie in the Swampland, and which are safely in the Landscape, a number of Swampland Conjectures have been put forward (for recent reviews see e.g.~\cite{Palti:2019pca, vanBeest:2021lhn,Grana:2021zvf}). Among them, the de Sitter Swampland Conjecture proposes that (meta)stable de Sitter vacua  are inconsistent with quantum gravity.  In terms of a string compactification's low energy EFT ingredients, it is supposed that the scalar potential of the string moduli (describing sizes, shapes and positions in the extra dimensions, and the string coupling) should satisfy \cite{Garg:2018reu,Ooguri:2018wrx}:
\be\label{sdSC}
\frac{\sqrt{\nabla^a V\nabla_a V}}{V} \geq \frac{c}{\Mp}  \qquad {\text{or}} \qquad 
  \frac{{\rm min} (\nabla^a \nabla_b V)}{V} \leq -\frac{c'}{\Mp^2}  \,,
\ee
where ``min()" denotes the minimal eigenvalue and $c$ and $c'$ $ {\mathcal O}(1)$ positive constants.  Whilst there exist physical arguments for these 
inequalities to hold in asymptotic regions of the moduli space \cite{Ooguri:2018wrx} -- where large moduli correspond to weak couplings in the corresponding perturbative expansions -- the conjecture speculates that it holds everywhere in moduli space.  This is not uncontroversial, but it is based on the empirical evidence  previously discussed, together with conceptual issues with observers in de Sitter space, such as how to define an S-matrix in this context \cite{Witten:2001kn, Banks:2012hx, Dvali:2018jhn}.

Conjecture \eqref{sdSC} rules out a metastable de Sitter vacuum as the explanation for DE, as we cannot have simultaneously $\nabla_a V =0$, $V>0$, and ${\rm eigenvalues}(\nabla^a \nabla_b V) >0$.  
The main alternative model for DE is slow-roll quintessence, and a priori it would be natural to expect that quintessence candidates are found amongst the string moduli. 
 Although  the conjecture is in tension with the simplest realisations of slow-roll inflation for the early Universe  -- the left-hand-sides of \eqref{sdSC} corresponding directly to the potential slow-roll parameters, $\epsilon_V$ and $\eta_V$, which need to be small in single-field slow-roll inflation (but see e.g.~\cite{Cicoli:2023njy} for potential counter-examples) -- it leaves some room to play within the context of late-time quintessence, depending on the values of $c$ and $c'$.  Indeed, whereas around 60 e-folds of inflation in the early Universe are required to solve the horizon problem, the late-time accelerated expansion has been occurring for less than one e-fold of expansion, suggesting a viable window in \eqref{sdSC}.  
It is important to note, however, that quintessence alternatives remain challenging to obtain in string theory, as they must also satisfy several phenomenological constraints, as well as address the cosmological constant problem. 
 
Keeping these caveats in mind, three distinct classes of simple slow-roll quintessence
 potentials then come to mind: plateaus, runaways, and hilltops.  Plateaus, corresponding to both first and second derivatives of the potential being small, are in tension with \eqref{sdSC}; moreover, they are as difficult to obtain from string theory as metastable dS vacua.  Runaway potentials are ubiquitous in string theory, and have both parametric control and a suppressed vacuum energy emerging at the asymptotics of field space.  However, generally these potentials   turn out to be too steep to source a slow-roll\footnote{Any potential can have some amount of accelerated expansion simply by tuning initial conditions such that the scalar starts by rolling up its potential.} accelerated expansion \cite{Rudelius:2021oaz, ValeixoBento:2020ujr}, consistently with the expectation from \eqref{sdSC} that the slow-roll parameters should be large.  Moreover,  the small window that could be consistent with \eqref{sdSC} and slow-roll quintessence does not agree with  observations. In fact, string compactifications generically lead to asymptotic exponential potentials, $V(\phi) = V_0 e^{-\lambda \phi}$, which satisfy the conjecture with $\lambda \gtrsim \sqrt{2}$.  Such potentials can source a transient late-time accelerated expansion that follows epochs of radiation and matter domination,  provided that $\lambda \lesssim \sqrt{3}$ \cite{Andriot:2024jsh}, but this parameter space turns out to be ruled out by the most recent cosmological data \cite{Bhattacharya:2024hep, Ramadan:2024kmn, Alestas:2024gxe}.   It is worth emphasising that these results demonstrate a powerful synergy between quantum gravity considerations and cosmological observations: by themselves quantum gravity would allow exponential quintessence with $\sqrt{2}\lesssim \lambda \lesssim \sqrt{3}$ and observations would allow exponential quintessence with $\lambda \lesssim 0.537$ \cite{Bhattacharya:2024hep}; taken together, exponential runaway quintessence is currently disfavoured.  It remains to consider the option of hilltop potentials: this is our
aim in this work. 

We focus on hilltop quintessence scenarios and explore the interplay between quantum gravity constraints and the most recent cosmological datasets.  The models we consider can be made consistent with Conjecture \eqref{sdSC} as
well as  other swampland conjectures; moreover, they are expected to arise naturally within  string theory. 
Specifically, we consider axion hilltops, hilltops for a saxion within 4D ${\cal N}=1$ supergravity, and a Higgs-like potential that, close to the hilltop, plays the role of a generic quadratic hilltop. 
In Section \ref{sec_theory}, we present these general models, motivating their initial conditions; in particular we provide a novel dynamical mechanism to set up the initial conditions for the axion hilltops.
After also discussing the connections of the models with string theory and particle physics, in Section \ref{sec_parametr} we review how they can be analysed in a unified way using an appropriate parameterisation of the quintessence equation of state, put forward by Dutta and Scherrer \cite{Dutta:2008qn} (considering  also its generalisation for more general thawing quintessence models by Chiba \cite{Chiba:2009sj}).

After presenting these general models, motivating their initial conditions, and discussing their connections with string theory and particle physics in Section \ref{sec_theory}, in Section \ref{sec_parametr} we review how they can be analysed in a unified way using an appropriate parameterisation of the quintessence equation of state, put forward by Dutta and Scherrer \cite{Dutta:2008qn} (considering  also its generalisation for more general thawing quintessence models by Chiba \cite{Chiba:2009sj}).  We  test this general parameterisation against our concrete scenarios and also show that it provides a useful analytical understanding of the degree of fine-tuning of initial conditions  necessary to be consistent with observations.

We are then ready, in Section \ref{sec_cosan}, to test our string-motivated hilltop models -- together with general hilltops using the Dutta-Scherrer parameterisation -- against a suite of recent cosmological data, from  CMB observations, galaxy surveys, and Type IA supernovae data.  We find the best-fit and mean values and bounds for the fundamental parameters in our models and investigate how these observational constraints stand against independent bounds from quantum gravity considerations. Furthermore, we identify which model is preferred by the data, comparing also to the fits of the $\Lambda$CDM model, the exponential runaway model, and the $w_0w_a$CDM model -- the latter corresponding to the alternative  Chevakkier-Polarski-Linder (CPL) parameterisation \cite{Chevallier:2000qy, Linder:2002et},  assuming  that the equation of state parameter evolves linearly with the scale factor.  We summarise our results in Section \ref{sec_conc}, where we also discuss the most important model-building challenges to be addressed in order
to extract the maximal information on quantum gravity scenarios from current and forthcoming data sets. A number of technical appendices follows. 

\section{Quintessence in string theory}
\label{sec_theory}

In this section we introduce the quintessence models whose cosmologies we study.  Importantly, these models are consistent with string theory expectations, but have different interpretations within string theory, and thus distinct associated microscopic parameters with specific interplays with particle physics.  After motivating and presenting the models 
 we analyze the associated cosmological equations that will be used in the following sections
to confront them with data\footnote{In the recent paper \cite{Arjona:2024dsr}, model-independent cosmological constraint on $c$ are explored, finding larger-than-one  values for this quantity. However,
the analysis of \cite{Arjona:2024dsr} does not include hilltops, and
the parameter $c'$ is not considered individually.   Specifically, \cite{Arjona:2024dsr} focuses on the combined quantity $\Gamma=-c'/c^2$ in the range $\Gamma>1$, which excludes hilltop scenarios (and exponentials, which have $\Gamma=1$).}.

\subsection{dS minima, plateaus, runaways, maxima and saddles vs the Swampland}

As  discussed in the introduction, the swampland conjectures on the properties of scalar potentials in string theory \eqref{sdSC} (and the difficulties in constructing controlled metastable dS string vacua that themselves motivated the conjectures)  suggest that the current acceleration of the universe cannot be attributed to a small positive cosmological constant.  The same swampland constraints \eqref{sdSC} are also in tension with a slow-roll quintessence characterised by a scalar potential with a plateau.  We have also reviewed how the conjecture \eqref{sdSC} suggests that runaways are typically too steep to source slow-roll accelerated expansion, and the small window that is allowed is ruled out by observations.

The task is to determine quintessence scenarios that address
the previous issues.
It is  interesting to note that dS maxima and saddles  generically satisfy the swampland constraints in \eqref{sdSC} and, moreover, seem to be  easier\footnote{\label{footnote:control} It should, however, be acknowledged that so far all explicit top-down constructions of dS maxima and saddles  involve some size moduli that are smaller than the string-length, implying that they might be spoiled by large $\alpha'$ corrections; see e.g. \cite{Parameswaran:2010ec, Andriot:2019wrs, Parameswaran:2024mrc}.  On the other hand, as we will discuss below, one can make rather general arguments for their existence.} to find in string constructions compared to dS minima, plateaus and slow-roll runaways.   In the following, we will use the term dS ``hilltops'' to refer to both dS maxima and dS saddles.  If one considers a modulus close to a dS hilltop, it can remain frozen by Hubble friction during the epochs of radiation and matter domination, acting as a small cosmological constant, until it only recently starts to slowly roll down its potential sourcing dynamical dark energy.  This provides a string-motivated, (thawing\footnote{Hilltop quintessence is an example of  {\em thawing} quintessence where $w_\phi$ starts close to -1 and then increases, as opposed to {\em freezing} quintessence where $w_\phi$ starts  above -1 and  decreases towards it.}) dynamical dark energy candidate\footnote{See~\cite{Giare:2024smz} for a recent work on interacting dark energy bounds after DESI.}, towards which -- tantalizingly -- recent data  seem to hint \cite{DES:2024tys,DESI:2024mwx,DESI:2024kob}\footnote{Of course, we have to wait for future more precise data to have a conclusive answer regarding the  dynamical behaviour of dark energy.}.
Nevertheless, we stress that the scenarios we discuss do not solve deep theoretical issues as the cosmological constant problem, since they do not directly explain the small size of dark energy energy density and how the Standard Model contributions are suppressed.

\subsection{String models of hilltop quintessence}
\label{sub_string}

We will consider three classes of string theory hilltop models, with the quintessence field descending from string moduli or matter fields.  Within string theory, two types of moduli -- both amongst the most generic predictions of the field content of string scenarios -- can be  identified as  hilltop quintessence candidates\footnote{Generally, more moduli could be rolling over their potential --  however we focus  on the simplest case of a single dynamical field.}:
\begin{enumerate}

\item {\bf Axion hilltops.}  Axions descend from the dimensional reduction of higher dimensional $p$-forms, giving rise to the so-called string axiverse\footnote{For early work on axions as quintessence in field theory see e.g. \cite{Frieman:1995pm, Choi:1999xn,Kim:2002tq,Svrcek:2006hf,Kaloper:2008qs,Panda:2010uq}.} \cite{Arvanitaki:2009fg}.   
Their scalar potential can be generated by non-perturbative effects such as instantons, with the leading contribution taking the form: 
\be\label{axionV}
    V(\theta) = V_0\left(1-\cos\left(\frac{\theta}{f}\right)\right) \,,
    \ee
where $f$ is the axion decay constant, or shift-symmetry breaking scale, and $V_0$ has an exponential suppression in the instanton action, $V_0 \sim M^4 e^{-S_{\rm inst}}$, with $M$ the scale of the instanton physics. Both $1/f$ and $S_{\rm inst}$ typically go as the saxionic superpartner for the axion, leading to $f S_{\rm inst} \sim x \Mp$ with $x$ an order number \cite{Svrcek:2006hf}.  Also, we have assumed (some resolution to the cosmological constant problem and) a Minkowski minimum at $\theta_{\rm min}=0$, whilst there is a dS maximum at $\theta_{\rm max}=\pi f$. 

Depending on the value of the decay constant, axion dark energy can occur either (i) close to the minimum of the potential at $\theta=0$, and up to its inflection point at $\theta=\pi f/2$, which requires $f>\Mp$, or (ii) near the hilltop at $\theta_{\rm max} =\pi f$, which allows for $f\lesssim\Mp$. We are interested in the latter case, since it allows for  values of the decay constant that are usually found in string theory constructions.  

Indeed, $f$ has been argued to be always $f \lesssim \mathcal{O}(1) \Mp$ by the so-called weak gravity conjecture \cite{Arkani-Hamed:2006emk}, which for axions implies that there must exist an instanton whose action satisfies\footnote{There is  a large literature that attempts to obtain super-Planckian decay constants, usually involving multiple axions for which there exists a generalised mutifield weak gravity conjecture (WGC).  The axion alignment mechnism \cite{Kim:2004rp} invokes two (or more) axions and a fine-tuning between their axion decay constants that produces a large effective axion decay constant.  In $N$-flation \cite{Dimopoulos:2005ac}, a large number $N$ of axions, each with axion decay constant $f$, lead to an effective axion decay constant $f_{\rm eff} = \sqrt{N} f$.  However, both axion alignment and $N$-flation require some extra model building to satisfy the multifield WGC, and in any case violate the strong WGC.  See e.g. \cite{Palti:2019pca} for a further discussion of the literature, including top-down model building attempts, where the challenges encountered may well seem consistent with a quantum gravity censorship of large axion decay constants.}:
\be \label{wgc}
S_{\rm inst} \lesssim \frac{\Mp}{f}\,,
\ee
In the strong version of the WGC this instanton must be the one with smallest action i.e. the leading effect. To keep control of the instanton expansion assumed in \eqref{axionV}, we require $S_{\rm inst} > 1$ and thus $f < \Mp$.  In the case that $S_{\rm inst} \gg 1$, the exponential suppression in $V_0$ naturally realises the necessary hierarchy between the dark energy potential and the leading-order potential that fixes the volume moduli.

As pseudo-scalars, axions can  evade  stringent  fifth force constraints even if they are extremely light. Furthermore, their approximate shift symmetries  restrict their allowed couplings and protect the axion mass and potential energy density, which are otherwise UV sensitive quantities.  An important open question is why initial conditions would be fine-tuned close to the hilltop. One possible mechanism\footnote{See \cite{Co:2018mho} for an alternative dynamical mechanism for the QCD axion.} to achieve this is if in the early Universe, the leading non-perturbative effects that are active stabilise the axion in a Minkowsi or adS minimum, and at some later time a further non-perturbative effect dynamically comes into play, turning this minimum into a (nearby) dS maximum (see Appendix \ref{app:suaxion} for a working example).

\item {\bf Saxion hilltops.}  Geometric string moduli, corresponding to sizes and shapes of the extra dimensions and positions within them, as well as the dilaton, also arise generically in string compactifications and are often associated with some small expansion parameter in the low-energy effective field theory description assumed.  They  also have potentials that include dS maxima and saddles; 
indeed, one can formulate rather general arguments for the existence of such hilltops.  Consider, for example, a compactification that stabilises all moduli in a regime of parametric or numerical control to a supersymmetric AdS vacuum, with one modulus lighter than the others. If the leading correction to the potential at asymptotic values of the light modulus is positive, then a dS maximum must exist to the right of the AdS minimum, and thus also under control \cite{Conlon:2018eyr}.  The setup becomes a priori more complex if there is more than one light modulus, since then the minimum may not be accompanied by an extremum in all directions.  It is therefore interesting to note that the recent explicit constructions  \cite{McAllister:2024lnt} of supersymmetric AdS minima in type IIB flux compactifications with many moduli  -- which are under numerical control\footnote{Note that -- even though the solutions in \cite{McAllister:2024lnt} include some two-cycle volumes that are small (c.f. footnote \ref{footnote:control}) -- it has been checked explicitly that e.g. the worldsheet instanton expansions are under control.  Moreover, whilst control of the dS minima (a.k.a.~``KKLT vacua'') in \cite{McAllister:2024lnt} is under question because the concrete examples have $g_s M \lesssim 1$, this regime is only of concern in the presence of a warped throat, in which case the supergravity expansion breaks down.  The warped throat is a necessary ingredient for the KKLT dS minimum, but not for the precursor supersymmetric AdS vacuum of interest here.} -- have  been found to be accompanied by dS maxima.

For concreteness, in the following, we  consider a specific, simple saxion hilltop model that is well-motivated from supergravity and was studied recently in \cite{Olguin-Trejo:2018zun}.  This model starts with a supersymmetric Minkowski setup with one flat direction, which is lifted by a leading-order supersymmetry-breaking non-perturbative contribution, with details given in Appendix \ref{app1}.  The corresponding scalar potential, expressed in terms of the canonical normalised field, is given by:
\be\label{sugraV}
    V(\phi) = V_0 \,e^{-\sqrt{2}\phi} \,e^{-2\alpha e^{\sqrt{2}\phi}} \lp -2+4\,\alpha^2 e^{2\sqrt{2}\phi} +4\,\alpha \,e^{\sqrt{2}\phi}\rp\,,
    \ee
with $\alpha$ a constant that depends on the type of non-perturbative effect in play.  For example, for gaugino condensation in a hidden $SU(N)$ gauge group (from wrapped D7-branes) in type IIB string models, we have $\alpha=2\pi/N$.  The scalar potential \eqref{sugraV} has a maximum at:
\be\label{eq:fimax}
    \phi_{\rm max} = \frac{1}{\sqrt{2}}\log\left(\frac{1}{\sqrt{2}\alpha}\right)\,,
    \ee
     which lies in a ``weak-coupling'' regime, say $\phi_{\rm max} \gtrsim 0.4$ for the canonically normalised field\footnote{This corresponds to the original field $ \varphi_{\rm max} \gtrsim 1.8$ (see  \eqref{eq:varphimax}).}, for around $\alpha \lesssim 0.4$,  or $N \gtrsim 16$.  Note that geometrical and topological constraints imply that $N$ cannot be arbitrarily large\footnote{E.g. for gaugino condensation from wrapped D7-branes in type IIB, \cite{Louis:2012nb} (see also \cite{Carta:2019rhx}) found that $N \lesssim {\cal O}(10)h^{1,1}$ with $h^{1,1}$ related to the number of size moduli of the compactification (more explicitly, $h^{1,1}$  is the Hodge number of the Calabi-Yau manifold counting for the number of K\"ahler moduli).}, but we can safely take, say, $N \lesssim \mathcal{O}(100)$.  On the other hand, as discussed in Appendix \ref{app1}, we can in any case expect at best numerical control of the expansion in the non-perturbative effects at the hilltop.

Although the exponential suppression in the saxion potential energy turns out to cancel at $\phi_{\rm max}$, its scale can match  the observed dark energy by being multiply exponentially suppressed in the vevs of the supersymmetrically stabilised moduli \cite{Parameswaran:2010ec}.  Saxions do not enjoy a shift symmetry like the axions, but constraints from 
time variation of fundamental constants, fifth forces and radiative corrections can potentially be avoided if the quintessence couples only indirectly, via gravity and with some further geometric suppression, to  the Standard Model and the supersymmetry breaking sector.  The fine-tuning of initial conditions could be explained e.g.~via high temperature effects \cite{Hardy:2019apu} or some other dynamics
\cite{Gomes:2023dat} turning the maximum into a transient minimum analogously to symmetry restoration in the Higgs potential; alternatively, anthropic arguments might be relevant, since without fine-tuning to the hilltop, the saxion would runaway to decompactification or decoupling and an unviable universe\footnote{These arguments do not work in the same straightforward way for axions.  E.g. the axion's remnant discrete shift symmetry implies that axion-matter couplings are such that finite temperatures, and other dynamical effects, typically only change the effective axion-mass and not the position of the minima in axion potentials (see e.g. \cite{Sikivie:2006ni, Hardy:2019apu}).  Also,  relaxing initial conditions away from the hilltop would lead to the axion rolling down to a minimum whose vacuum energy is of similar magnitude to that at the maximum and hence equally anthropically viable.}.

\end{enumerate}

In what follows, we explore hilltop quintessence using the two concrete moduli examples above --  the axion hilltop and the supergravity saxion hilltop -- together with a more generic Higgs-like  quadratic   hilltop, which can approximate any quadratic hilltop near the top, and which might descend from a stringy saxion modulus, stringy axion modulus, or a stringy matter field:
\begin{enumerate}\setcounter{enumi}{2} 
    \item {\bf Higgs-like hilltops.}   The dynamics from any quadratic hilltop potential can be approximated by using a Higgs-like potential
    \be\label{eq:FTV}
    V(\phi) = V_0\lp 1-\lb\frac{\phi}{\phi_0}\rb^2\rp^2\,.
    \ee
      This potential is bounded from below by having a quartic term compared to the more standard quadratic field theory hilltops. This -- and any other terms that might appear in the Taylor expansion of the hilltop potential about its maximum -- will not affect the dynamics for as long as the field stays sufficiently close to its hilltop. 

      Should the field $\phi$ explore a significant part of its field-range, the swampland distance conjecture \cite{Ooguri:2006in} would limit $\phi_0 \lesssim \mathcal{O}(1) \Mp$.  In hilltop quintessence scenarios, $\phi$ remains frozen for most of the cosmological history, allowing this constraint to be relaxed.  
\end{enumerate}
We focus on the cosmological aspects of the hilltop quintessence models, from the cosmological evolution and comparison with cosmological observations in the CMB, galaxy surveys and type IA supernovae catalogues, to a discussion on the initial conditions and implications for inflation and reheating.  We aim to ascertain to what extent theoretical and observational constraints might favour a specific  model of hilltop quintessence, and what insights observational constraints give into the microscopic parameters of stringy hilltop quintessence models.

\subsection{Cosmological equations for hilltop quintessence}

We now set up the equations of motion that describe the background cosmological evolution for the models of interest.  We consider a universe whose dark energy (DE) component is described microscopically by one of the hilltop quintessence fields introduced above, with a canonical kinetic term and a scalar potential functional $V(\phi)$, minimally coupled to gravity.  We further include radiation and (dark) matter and -- though we do not consider the detailed string theory model building required to achieve it -- we assume that they are decoupled from the quintessence field. Given theoretical expectations and observational prospects, we also allow for non-zero curvature of the 3D space slices at this stage\footnote{For completeness, in Appendix \ref{app2} we collect the cosmological evolution including curvature.}. 

The 4D FLRW metric  with arbitrary curvature given is given by:
\be\label{metric}
ds^2 = -dt^2 +a^2(t) \left(\frac{dr^2}{1-kr^2}+r^2\left[d\theta^2+\sin^2{\theta}d\varphi^2\right]\right) \,,
\ee
where $k=0,\pm1$ denotes the curvature of the 3D slices. 
The energy momentum tensor is described by a set of perfect fluids describing the radiation, matter, quintessence, and effective ``curvature fluid'' components. 
The energy density and pressure, $\rho_i,p_i$, for these components are related by their {\em equation of state parameter}, $w_i$, as:
\be
p_i=w_i\rho_i\,,
\ee
where $i=r, m, \phi, k$ runs over radiation, matter, quintessence, and curvature; $w_r =\frac{1}{3}$ and $w_m=0$, whilst for the scalar:
\be\label{eq:rhofi}
\rho_\phi = \frac{\dot\phi^2}{2} + V(\phi) \,,\qquad p_\phi = \frac{\dot\phi^2}{2} - V(\phi)\,, \qquad w_\phi = \frac{p_\phi}{\rho_\phi},
\ee
and for the curvature component:
\be\label{eq:rhok}
\rho_k = -\frac{3\,k}{a^2}\,,  \qquad  p_k = \frac{k}{a^2} \,, \qquad w_k = -\frac{1}{3} \,.
\ee

We can now write down the cosmological equations of motion for this system, which are given by (we set $\kappa= 8 \pi G_N =1$ for now):
\begin{subequations}\label{eq:eoms}
 \begin{align}
      H^2 &= \frac{\rho_{\rm eff}}{3}\,,
      \label{eq:Fried}\\
    \frac{ \ddot a}{a} &
    =-\frac{\rho_{\rm eff}}{6}(1+3w_{\rm eff})\,,
    \label{eq:addot}\\
    \ddot \phi &= -3H\dot \phi -V_\phi \,.  \label{eq:phi}
 \end{align}
\end{subequations}
In the last equation, $V_\phi\equiv \del_{\phi} V$.  Moreover,  we  defined
\be\label{eq:effrhop}
\rho_{\rm eff} = \sum_i \rho_i\,, \qquad p_{\rm eff} = \sum_i p_i\,, \qquad p_{\rm eff}=w_{\rm eff}\, \rho_{\rm eff}\,.
\ee
 From this definition it is clear that 
\be\label{eq:weff}
w_{\rm eff} = \sum_i w_i\Omega_i\,, 
\ee
where 
\be
\Omega_i=\frac{\rho_i}{3H^2}\,.
\ee
Moreover, from \eqref{eq:addot}, we learn that  acceleration requires $w_{\rm eff}<-1/3$.

\section{Parameterisation of the equation of state for hilltop models}
\label{sec_parametr}
In all hilltop quintessence models, the quintessence field is initially frozen by Hubble friction close to the hilltop and starts to slowly roll, as the Universe expands and Hubble friction falls, in recent times.  All hilltop quintessence potentials are therefore well-approximated for the full cosmological history by their behaviour close to the hilltop, i.e.~their Taylor expansion around the maximum up to second order.  It is therefore not surprising that all hilltop models can be described in a universal way.  
In fact, Dutta and Scherrer have derived in \cite{Dutta:2008qn} a parameterisation of the equation of state parameter  for hilltop quintessence models, and Chiba has shown  in \cite{Chiba:2009sj} that this parameterisation can actually be extended to more general thawing quintessence models.

In this section, we first outline the derivation of the Dutta-Scherrer(-Chiba) (DS(Ch)) parameterisation and then test how it fairs for the specific hilltop models in our focus, showing that it performs much better than the more commonly used CPL parametersation \cite{Chevallier:2000qy, Linder:2002et}.  For comparison, we also test the DSCh parameterisation against a quintessence model without hilltop, specifically  with an exponential potential.  Finally, we use the DS parameterisation to obtain a bound on the initial displacement from the hilltop, which, as expected, depends on the curvature at the hilltop.  As we will show, this subsequently leads to a bound on the scale of inflation or reheating.

\subsection{Derivation of the Dutta-Scherrer(-Chiba) (DSCh) parameterisation} 
Dutta and Scherrer obtained their parameterisation of the equation of state parameter for hilltop quintessence by computing  the general solution to the scalar field in a flat universe whose dynamics is dominated by the scalar and matter.  
 We now outline the calculation of DS \cite{Dutta:2008qn}, pointing out along the way how the derivation also applies in the presence of curvature, so long as the curvature is subdominant, as it is in our observed universe.   As already mentioned, the beauty of the DS parameterisation is that it is analytically justified for  generic hilltop potentials, which can all be approximated by the same Taylor expansion around the maximum as:
\be\label{eq:Vexp}
V(\phi) \approx V(\phi_{\rm max}) + \frac12 V''(\phi_{\rm max})(\phi-\phi_{\rm max})^2 \,.
\ee

To begin, we  consider the scalar field equation, \eqref{eq:phi}, in the background of matter and the hilltop potential energy, which corresponds to an effective cosmological constant $\Lambda = V(\phi_{\rm max})$.  That is, we  assume that the rolling of the scalar field away from its hilltop hardly affects the overall background expansion of the Universe, and moreover, that radiation and curvature are negligible during the epochs of interest.  Note that the energy density from the curvature grows more slowly than that from radiation and matter as one tracks backwards in time, so if the curvature is subdominant today -- as it is -- then it was subdominant throughout the history of the universe.  Given our assumed background cosmology, which is effectively a flat $\Lambda$CDM (i.e.~neglecting the subdominant radiation, curvature and time-varying quintessence), the expansion as a function of time is given by:
 \be\label{eq:LCDM}
 \frac{a(t)}{a_0} =\left[\frac{1-\Omega_{\Lambda,0}}{\Omega_{\Lambda,0}}\right]^{1/3} \sinh^{2/3}(t/t_{\Lambda}) \,,
 \ee
 where $a_0$ is the scale-factor today, we denoted the present-day density parameter from the hilltop potential energy as $\Omega_{\Lambda,0}$, and  defined 
 \be
 t_{\Lambda} \equiv \frac{2}{3 H_0 \sqrt{\Omega_{\Lambda,0}}} =\frac{2}{\sqrt{3V_{\rm max}}}\,,
 \ee
with $V_{\rm max} \equiv V(\phi_{\rm max})$.
Let us now define a new variable $u(t)$ as follows \cite{Dutta:2008qn}:
\be\label{eq:udef}
u(t)=(\phi - \phi_{\rm max})a^{3/2}(t) \,.
\ee
In terms of $u$, the equation for $\phi$ \eqref{eq:phi} becomes 
\be\label{eq:ueq1}
\ddot u -\frac32 u \left[\frac{\ddot a}{a} +\frac12 H^2\right] +a^{3/2} V_\phi=0\,.
\ee
Using \eqref{eq:Fried} and \eqref{eq:addot}, this becomes
\be\label{eq:ueq2}
\ddot u +\frac34 u \,p_{\rm eff} +a^{3/2} V_\phi=0\,,
\ee
where remember that  we are assuming  that $p_{\rm eff}$ includes only the contribution from the hilltop maximum (with $w_\Lambda = -1$) and matter (with $w_m=0$) and therefore:
\be
p_{\rm eff} \simeq -V_{\rm max}\,.
\ee
Recalling the expansion \eqref{eq:Vexp}, and writing $V_{\rm max}'' \equiv V''(\phi_{\rm max})$, we have $V_\phi = V_{\rm max}'' (\phi-\phi_{\rm max})$.  Therefore, \eqref{eq:ueq2} can be reduced to:
\be\label{eq:ueq3}
\ddot u +u\left[V_{\rm max}''-\frac34 \,V_{\rm max} \right]=0\,.
\ee
 Defining 
 \be\label{kdef}
 k_V\equiv \sqrt{\frac34 V_{\rm max} -V_{\rm max}''} ~ ~\,,
 \ee
 the general solution to \eqref{eq:ueq3} is given by 
 \be\label{eq:usol}
 u(t) = A\,\sinh(k_Vt) + B \cosh(k_Vt)\,.
 \ee
 
 To fix the integration constants,  consider an  initial, finite value for $\phi$ at $t=0$:
 \be\label{eq:Dphi_i}
 \phi(0) \equiv \phi_i \equiv \phi_{\rm max} + \Delta\phi_i\,.
 \ee
Then \eqref{eq:udef} together with \eqref{eq:LCDM} requires $B=0$ in \eqref{eq:usol}, and also fixes the value of $A$ in terms of $\Delta \phi_i$, giving the final solution for $\phi=\phi_{\rm max} + u/a^{3/2}$:
 \be\label{eq:phisol}
 \phi = \phi_{\rm max} + \frac{\Delta\phi_i}{K }\frac{\sinh(k_V t)}{\sinh(t/t_\Lambda)}\,,
 \ee
 where we defined 
 \be\label{eq:Kdef}
 K\equiv k_V t_\Lambda.
 \ee

 The next step is to find an expression for the equation of state parameter associated to $\phi$ as a function of $a$, $w_\phi(a)$. First notice that 
 \be
 1+w_\phi = \frac{\dot\phi^2}{\rho_{\phi}}  \simeq \frac{\dot\phi^2}{V_{\rm max}}\,,
 \ee
where we used the approximation that $\rho_\phi\simeq V_{\rm max}$. 
Normalising this expression to the present-day value of $w_\phi$,  denoted by  $w_{0}$, and using the solutions for $\phi$ and the scale factor $a$, eqs.~\eqref{eq:phisol} and \eqref{eq:LCDM},  one arrives at the final expression as in \cite{Dutta:2008qn}:
\be\label{eq:wparam}
\frac{1+w_{\phi}(a)}{1+w_0} =
\left(\frac{a}{a_0}\right)^{3(K-1)}
\!\left[\frac{(K-F(a))(1+F(a))^K+(K+F(a))(F(a)-1)^K}{(K-F_0)(1+F_0)^K+(K+F_0)(F_0-1)^K}\right]^2\,,
\ee
where, approximating $\Omega_{\Lambda,0} \simeq \Omega_{\phi,0}$:
\begin{subequations}
\begin{align}
    K = &
    \sqrt{1-\frac43\frac{V''_{\rm max}}{V_{\rm max}}}\,, \label{eq:K}\\
    F(a) \equiv & \sqrt{1+\left(\frac{a}{a_0}\right)^{-3} \left(\frac{1-\Omega_{\phi,0}}{\Omega_{\phi,0}}\right)}\,,
    \qquad F_0=F(a_0)\,.
    \end{align}
\end{subequations}
We see that the parameterisation involves two free parameters:  $K$, which depends on the curvature of the potential around the hilltop as in \eqref{eq:K}; and $w_0$, which depends on $K$ and the initial displacement of the scalar field from its maximum, $\Delta\phi_i$, as:  
 \be\label{eq:phi_ini}
1+w_0=\frac{3}{16}\frac{\Delta\phi_i^2}{K^2}\frac{(1-\Omega_{\phi,0})}{\Omega_{\phi,0}}\left[\left(K-F_0\right)\left(\frac{1+\sqrt{\Omega_{\phi,0}}}{\sqrt{1-\Omega_{\phi,0}}}\right)^K +
\left(K + F_0\right)\left(\frac{1-\sqrt{\Omega_{\phi,0}}}{\sqrt{1-\Omega_{\phi,0}}}\right)^K
\right]^2 \,.
 \ee
 \smallskip

Chiba's generalization of (\ref{eq:wparam}) \cite{Chiba:2009sj} purports to extend the validity of the parameterisation beyond hilltops to general thawing quintessence models by allowing the field to start at some arbitrary initial value, $\phi_i$, and keeping all terms in the Taylor expansion of the potential around $V(\phi_i)$ up to second order (c.f.~(\ref{eq:Vexp})). Formally, this parameterisation ends up coinciding with (\ref{eq:wparam}), but with $K$ in (\ref{eq:K}) defined via the initial values of $V''_i$ and $V_i$, rather than their values at the maximum.  On the other hand, it is not obvious why -- if both first and second order terms in the Taylor expansion are significant -- third order terms and beyond can be neglected.  

It is  interesting to compare the DSCh  parameterisation with the commonly used  linear CPL parameterisation \cite{Chevallier:2000qy,Linder:2002et}, with parameters $w_0$ and $w_a$:
\be\label{eq:CPL}
w_\phi(a) = w_0+(1-a)w_a\,,
\ee
which -- though it lacks the analytical justification of the DS parameterisation for hilltops -- is based on a Taylor expansion of the equation of state parameter itself, where the leading term is the linear one.  In the following, we  test the DS(Ch) (\ref{eq:wparam})  and CPL (\ref{eq:CPL}) parameterisations against both hilltop models and the exponential runaway potential, using the modification by Chiba \cite{Chiba:2009sj} for the latter. 
A comparison of different phenomenological parameterisations was performed in \cite{Pantazis:2016nky} for a particular hilltop model with $V(\phi) = V_0\, e^{-c\phi}(1+\alpha\phi)$, while the DSCh parameterisation was analysed against recent data in \cite{Gialamas:2024lyw}.

\subsection{Testing the DS parameterisation for  hilltop quintessence models }
\label{sec:paramvsmodels}

In this section, we test the DS \cite{Dutta:2008qn} parameterisation against the explicit hilltop models presented in Section \ref{sec_theory}. For comparison, in the next subsection \ref{subsec:exp}, we test the parameterisation also against the exponential dark energy model, using the Chiba \cite{Chiba:2009sj}  generalisation.

We start by checking  how the parameter $K$ in the DS parameterisation is related to the more fundamental parameters in the  different models that we are interested in analysing: 
\begin{enumerate}
    \item {\bf Axion hilltops}. The scalar potential is given by \eqref{axionV}  
  and for hilltop quintessence, we are interested in the case $f\lesssim 1$, which is also consistent with typical values found in string theory and suggested  by swampland constraints \cite{Arkani-Hamed:2006emk}. For this model, the parameter $K$ is given by:
\be
K_{\rm ax} = \sqrt{1+\frac{2}{3f^2}}\,.\label{eq:Kax}
\ee
Thus for $f\lesssim 1$, $K_{\rm ax}\gtrsim \sqrt{5/3}\sim 1.3$.
In Figure \ref{fig:cos} we compare the true evolution of the equation of state parameter obtained using {\tt CAMB} with the DS and CPL parameterisations, 
for the best-fit value of the  decay constant using DESI year one data plus
Union3 supernova data, $f=0.15$ (see Table \ref{tab:Axion_table_full} in Appendix~\ref{app:fullmcmc}), and for $f=0.5$. As we commented before, since the spatial curvature $k$ is subdominant throughout the cosmological evolution, adding a small non-zero $k$ does not change the results. We collect the evolution with non-zero spatial curvature in Appendix \ref{app2}. 

As we can see from the Figure \ref{fig:cos}, the DS parameterisation works very well through the full cosmological evolution, well beyond the reach of current and near-future Dark Energy surveys; indeed, the derivation of the DSCh parameterisation suggests that it should work as soon as radiation is negligible (and recall that $z_{eq}\approx 3400$).  In the figures, the evolution starts  in the matter domination epoch from  $z_m=3000$. For the CPL parameterisation we fit the linear behaviour to obtain suitable values of $w_0, w_a$. It is  very clear that the linear parameterisation \eqref{eq:CPL} is not appropriate for axion hilltop quintessence throughout the evolution; rather it works only for small red-shifts as $f$ decreases.  We will see the same pattern also for the other hilltop potentials below.  

Finally, in Figure \ref{fig:cos2}, we compare the evolution of the dark energy equation of state for different initial conditions as indicated in the plot. 

\begin{figure}[H]
    \centering
    \includegraphics[width=0.95\linewidth]{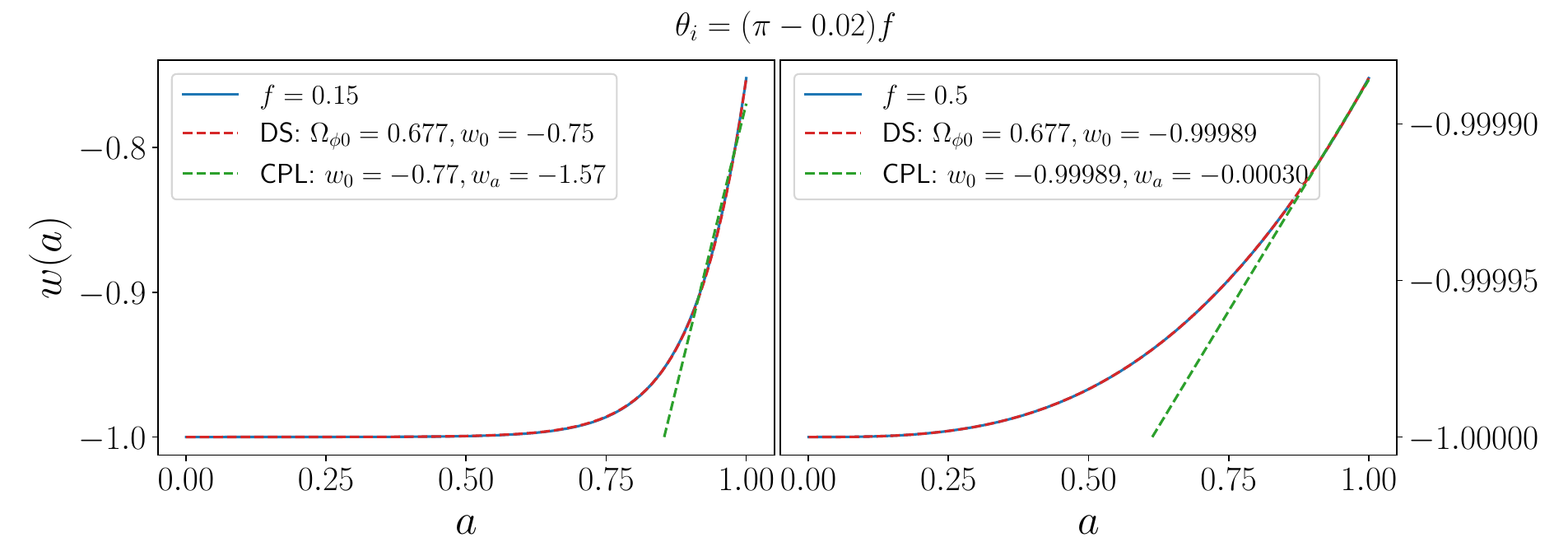}
    \caption{Evolution of the equation of state parameter for the axion hilltop for different values of $f$, and its comparison to the DS  \eqref{eq:wparam} and CPL \eqref{eq:CPL} parameterisations. The initial value for $\theta$ is given at the top of the plots and for the DS parameterisation we used the values of $\Omega_{\phi 0}$ and $ w_0$ as indicated to the right, as obtained from the evolution with {\tt CAMB}. The CPL parameters are obtained by fitting the linear behaviour  between $a=0.9$ and $a=1$. }
    \label{fig:cos}
\end{figure}

\begin{figure}[H]
    \centering
    \includegraphics[width=0.65\linewidth]{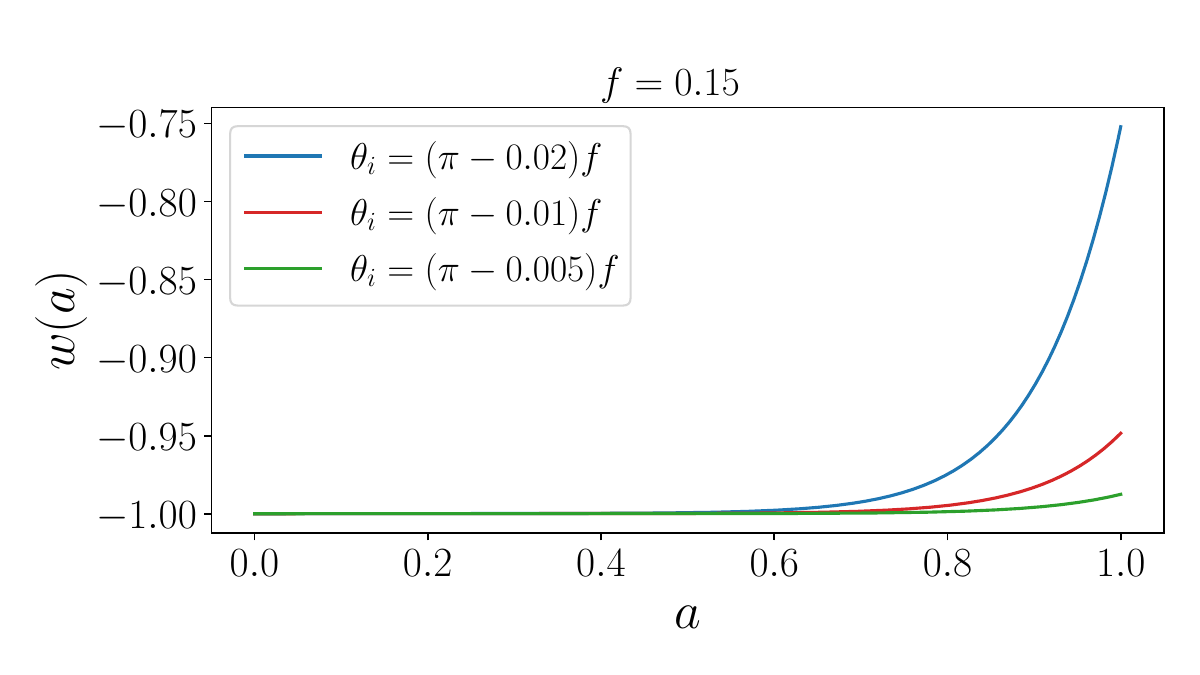}
    \caption{Evolution of the equation of state parameter for the axion hilltop for different initial conditions $\theta_i$.}
    \label{fig:cos2}
\end{figure}


    \item {\bf Saxion hilltop}.  The potential of this model is given by \eqref{sugraV} \cite{Olguin-Trejo:2018zun}, where 
    typical values for $\alpha$  are $\frac{2\pi}{N}$ with, recall, $N$ an integer that we assume to be $N \lesssim \mathcal{O}(100)$. 
    Interestingly, for this model,  $K$  is independent of the potential parameters!  Indeed,  
    \be
    K_{\rm sugra}= \sqrt{\frac{19+8\sqrt{2}}{3}}\simeq 3.179\,,\label{eq:Ksugra}
    \ee
and thus the curvature at the maximum is independent of $\alpha$ and the evolution will mostly be dependent on $\phi_i$. To better understand the constraints on $\alpha$ in this case, it is useful to write the potential as follows:
\be\label{eq:Vsugraexp}
V=\frac{e^{-\sqrt{2}}}{3}\,V_0\,\alpha \lb a_0 - a_2 \,\kappa^2\phi_{\rm max}^2\lp\frac{\phi}{\phi_{\rm max} }-1\rp^2 +a_3 \,\kappa^3\phi_{\rm max}^3\lp\frac{\phi}{\phi_{\rm max}}-1\rp^3 + \dots\rb \,,
\ee
where we restored  Planck units $\kappa=1/M_{\rm Pl}$, $a_n$ are numerical constants independent of $\alpha$ given by  $a_0=12$, $a_2= 12 (2 + \sqrt{2})$ and $a_3=32$, and recall that $\phi_{\rm max}$ is given in terms of $\alpha$  by \eqref{eq:fimax}. Note that  $\phi_{\rm max}$ can be positive or negative, depending on the value of $\alpha$ (see eq.~\eqref{eq:fimax}). In particular for $\alpha\geq 1/\sqrt{2}$, $\phi_{\rm max}\leq 0$.  On the other hand,  $\phi_{\rm max}$ becomes super-Planckian for $\alpha\lesssim 0.17\sim \frac{2\pi}{37}$. As we mentioned before, $\alpha$ is also constrained by ensuring theoretical control to $\alpha\lesssim 0.4$. 
Therefore in the next section we will focus on  $\alpha\in \lp\frac{2\pi}{32},\frac{2\pi}{9}\rp$. 

From the expansion around the maximum \eqref{eq:Vsugraexp}, we also see that 
contrary to the axion and field theory Higgs-like model, there is a  non-zero cubic contribution. Thus, for $\phi/\phi_{\rm max}<1$, the cubic term changes sign, as the potential becomes steeper (and unbounded) on the left hand side. (On the other hand, the axion and Higgs potentials have the same curvature to either side). Due to the cubic contribution, we expect the DS parameterisation to be a little less accurate in this case, compared to the axion and Higgs. 

In Figure \ref{fig:sugra} we compare the evolution of the equation of state with the DS and CPL parameterisations, using the best-fit value for the parameter $\alpha$ obtained from DESI year one data plus Union3 supernova data i.e.~$\alpha=0.37\sim \frac{2\pi}{17}$ (see Table \ref{tab:Sugra_table_full} in Appendix~\ref{app:fullmcmc}). From the figure we see that the DS parameterisation does slightly worse here compared to the Axion  (see Figure~\ref{fig:cos}) and Higgs (see Figure~\ref{fig:FT}) cases, through the full cosmological evolution, whereas the linear parameterisation breaks down rather quickly.

In Figure \ref{fig:sugra2}, we compare the evolution of the dark energy equation of state for the saxion model, for different initial conditions. 

\begin{figure}[H]
    \centering
    \includegraphics[width=0.65\linewidth]{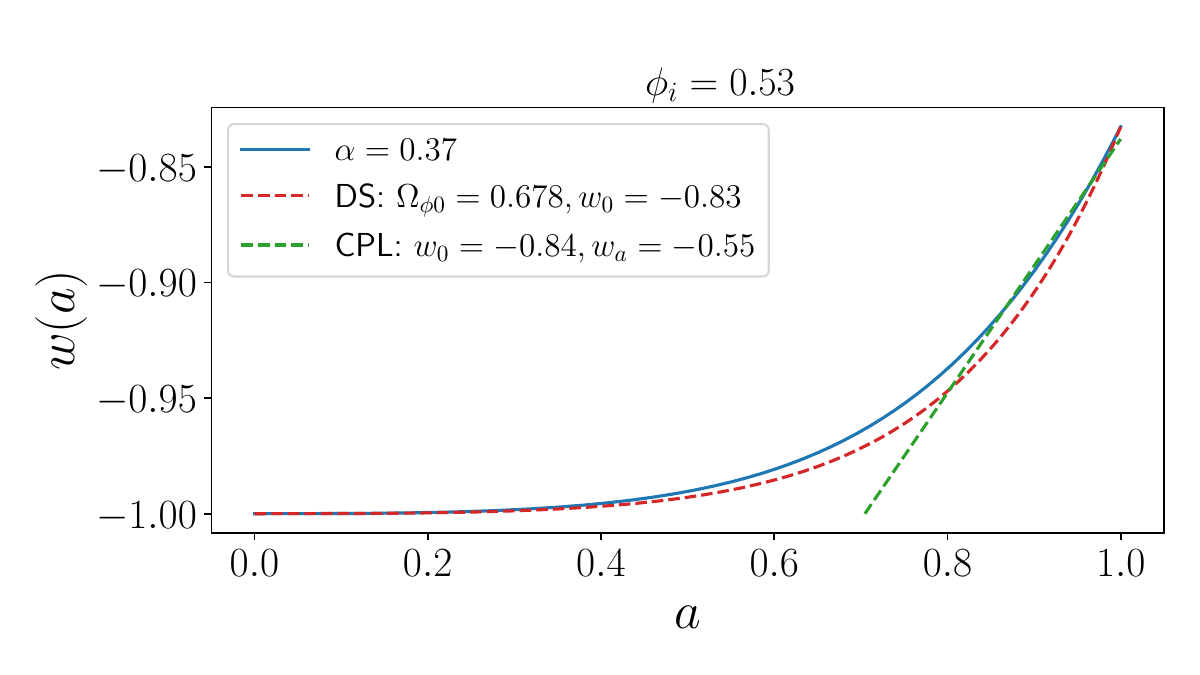}
    \caption{Evolution of the equation of state parameter for the supergravity hilltop  and its comparison to the DS \eqref{eq:wparam} and CPL \eqref{eq:CPL} parameterisations. The initial value for the saxion is given at the top of the plot. For the DS   parameterisation we used the values of $\Omega_{\phi 0}$, $w_0$ obtained from the evolution with {\tt CAMB} as indicated to the right, while we fit the linear behaviour  between $a=0.9$ and $a=1$ to obtain $(w_0, w_a)$.} 
    \label{fig:sugra}
\end{figure}

\begin{figure}[H]
    \centering
    \includegraphics[width=0.65\linewidth]{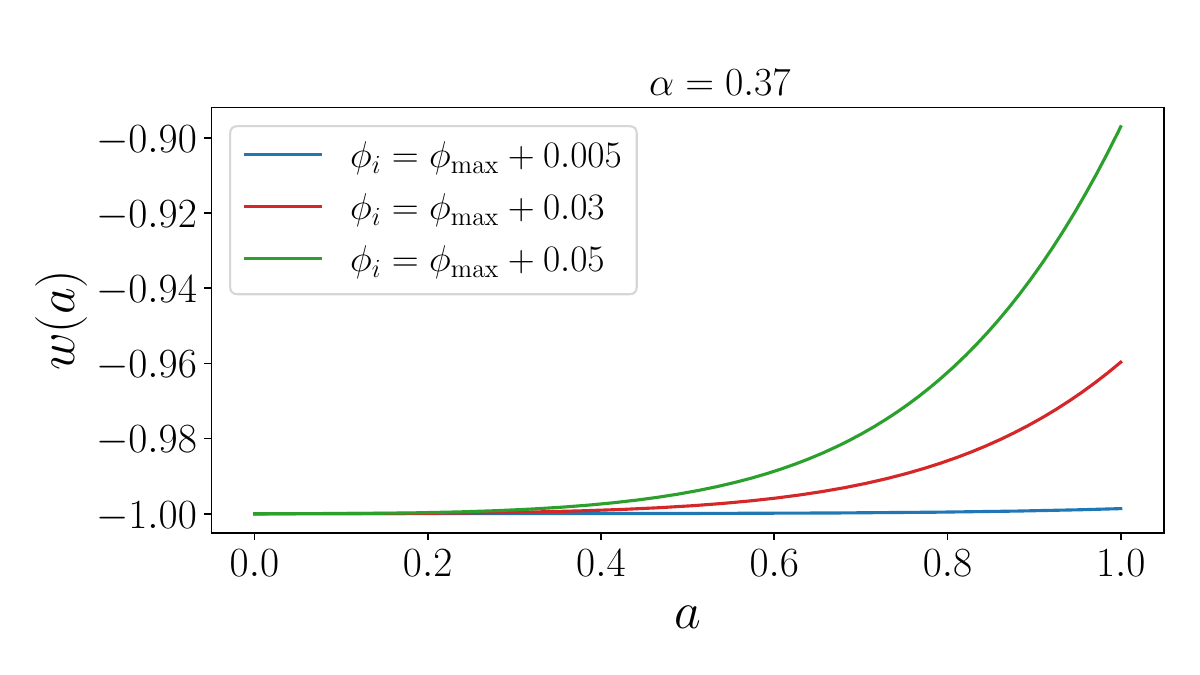}
    \caption{Evolution of the equation of state parameter for the supergravity hilltop for different initial conditions $\phi_i$.} 
    \label{fig:sugra2}
\end{figure}

    \item {\bf Higgs-like hilltop}.  The potential for this model is given by \eqref{eq:FTV}, where recall that in order to avoid issues with the numerical and cosmological analysis, we completed the quadratic hilltop to a  Higgs-like potential.  For this potential,     $K$ is given by
    \be
    K_{\rm Higgs} =\sqrt{1 + \frac{16}{3 \phi_0^2}}\,.\label{eq:Kft}
\ee
If we expect  $\phi_0$ to be less than or at most one (recall we are using Planck units), $\phi_0\lesssim 1$, then $K_{\rm Higgs}\gtrsim \sqrt{19/3}\sim 2.5$.

In Figure \ref{fig:FT} we compare the evolution of the equation of state with the DS and CPL parameterisations, for the best-fit value of $\phi_0=0.69$ using from DESI year one data plus
Union3 supernova data (see Table \ref{tab:Higgs_table_full} in Appendix~\ref{app:fullmcmc}) and for $\phi_0=1.3$.  Again, we learn that  the DS parameterisation works very well through the full cosmological evolution also in this example, whereas the CLP parameterisation works only for smaller redshifts. 

Finally in Figure \ref{fig:FT2} we compare the evolution of the equation of state for the field theory model for different values of $\phi_i$.

\begin{figure}[H]
    \centering
    
    \includegraphics[width=0.95\linewidth]{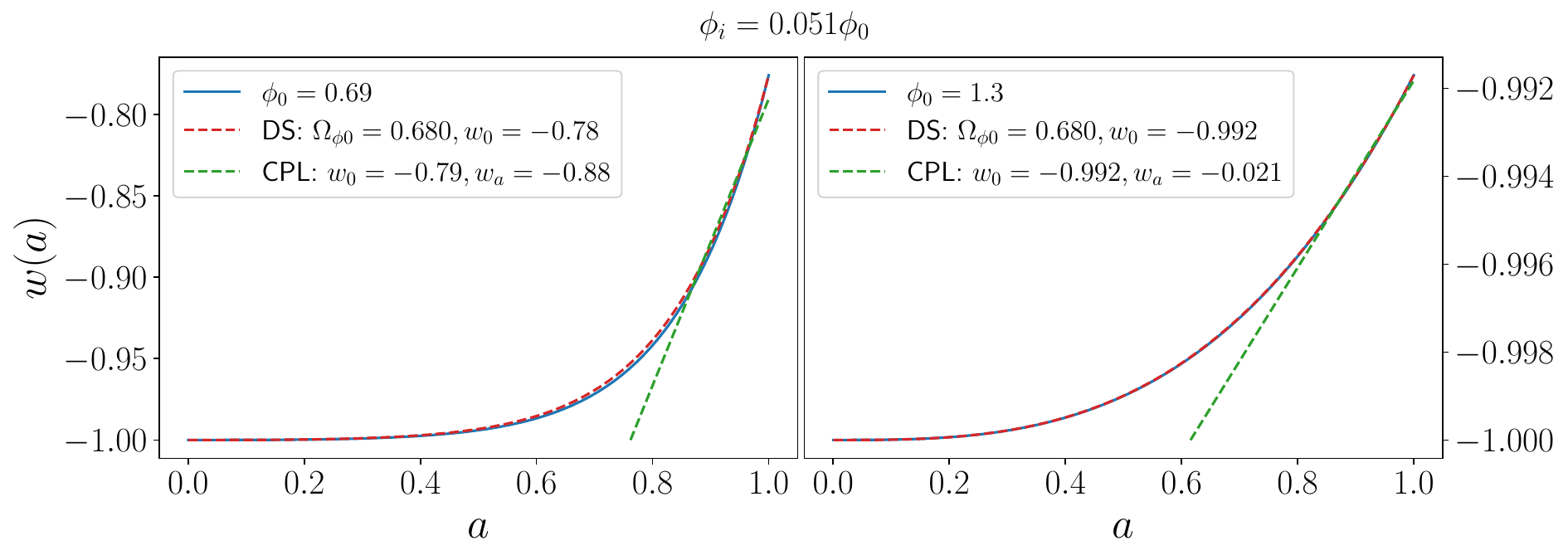}
    \caption{Evolution of the equation of state parameter for the Higgs-like hilltop \eqref{eq:FTV} and its comparison to the DS  \eqref{eq:wparam} and \eqref{eq:CPL} parameterisations. For DS we used $\Omega_{\phi 0}$, $w_0$ as obtained from the evolution with {\tt CAMB}, while we fitted the linear behaviour  between $a=0.9$ and $a=1$ for the CPL to obtain $(w_0, w_a)$ as indicated to the right. The  initial value for the scalar field is indicated at the top of the plot. }
    \label{fig:FT}
\end{figure}

\begin{figure}[H]
    \centering
    \includegraphics[width=0.65\linewidth]{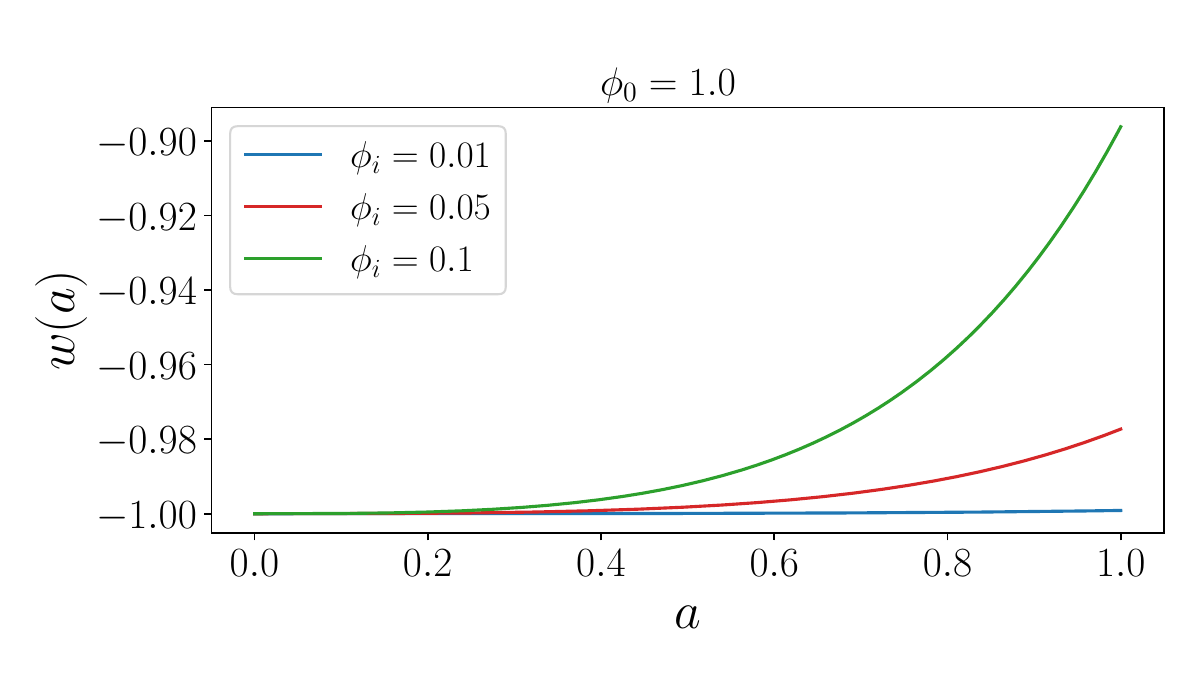}
    \caption{Evolution of the equation of state parameter for the Higgs-like hilltop for different initial conditions, $\phi_i$.  }
    \label{fig:FT2}
\end{figure}

\end{enumerate}

\subsection{Comparison: testing DSCh parameterisation in exponential quintessence}\label{subsec:exp}

\begin{figure}[H]
    \centering
    \includegraphics[width=0.95\linewidth]{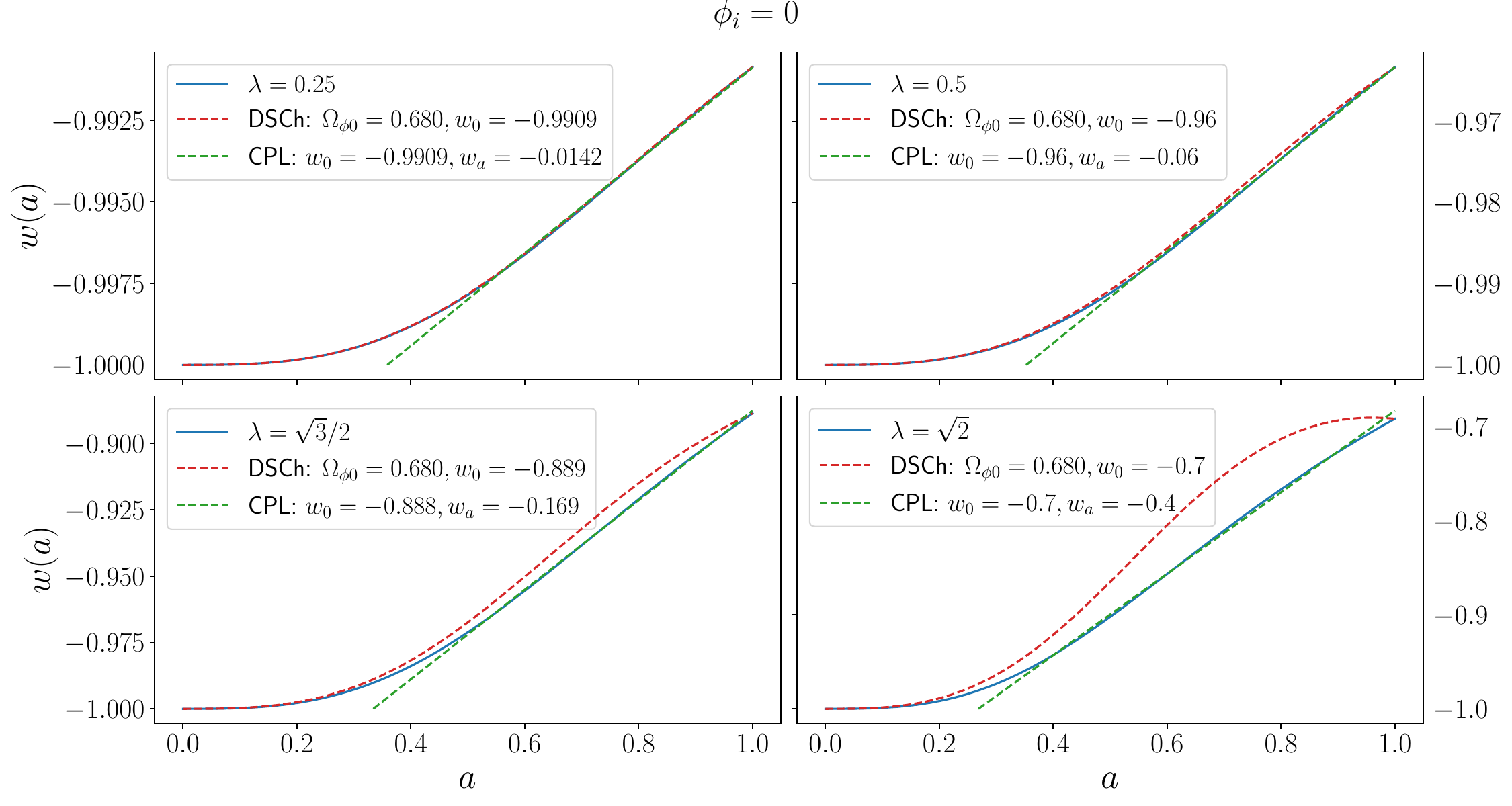}
    \caption{Evolution of the equation of state parameter for the exponential potential \eqref{eq:Vexp} and its comparison to the DSCh and CPL parameterisations. For the latter we used $\Omega_{\phi 0}$, $w_0$  obtained from the evolution with {\tt CAMB}, while for CPL we fitted the linear behaviour between $a=0.5$ and $a=1$ to obtain the parameters, indicated to the right.}
    \label{fig:exp}
\end{figure}

For comparison, having verified the success of the DS parameterisation for various explicit hilltop models, we now test how its generalisation by Chiba fares with a potential that does not have a maximum. We do this using an exponential potential, whose cosmology has recently been studied in \cite{Akrami:2018ylq,Bhattacharya:2024hep,Ramadan:2024kmn}:
\be
V=V_0 \, e^{-\lambda \phi}\,.
\ee
Recall that the DSCh parameterisation corresponds to \eqref{eq:wparam} where $V_{\max}, V''_{\rm max}$ in $K$ are  replaced by $V_i=V(\phi_i), V_i''=V''(\phi_i)$. Though in general $K$ will depend on $\phi_i$, for the exponential case, it is independent of it, becoming purely dependent on $\lambda$: 
\be
K_{\rm exp}= \sqrt{1 - \frac{4 \lambda^2}{3}}\,.
\ee
Note that $K_{\rm exp}^2$ can be negative, giving rise to oscillatory behaviour in the parameterisation \cite{Chiba:2009sj}. In this case, one should replace $K\to i\tilde K$ with $\tilde K=\sqrt{4V''_i/3V_i -1}$ in \eqref{eq:wparam}.
Interestingly, $K_{\rm exp}^2<0$ for $\lambda>\sqrt{3}/2\sim 0.866$. 
In Figure \ref{fig:exp} we compare the evolution of the equation of state with the DSCh and CPL parameterisations  for different values of $\lambda$. As we can see from the comparison, the DS parameterisation works rather well for $K^2_{\exp}>0$, but it does not do well for  $K^2_{\exp}<0$. For $K^2_{\exp}>0$, the DSCh parameterisation works out to larger redshifts than the linear CPL parameterisation, but for $K^2_{\exp}<0$ the CPL parameterisation does better.  

\subsection{Bounds on -- and from -- the initial conditions}\label{sec:bounds}

\begin{figure}[H]
    \centering
    \includegraphics[width=0.7\linewidth]{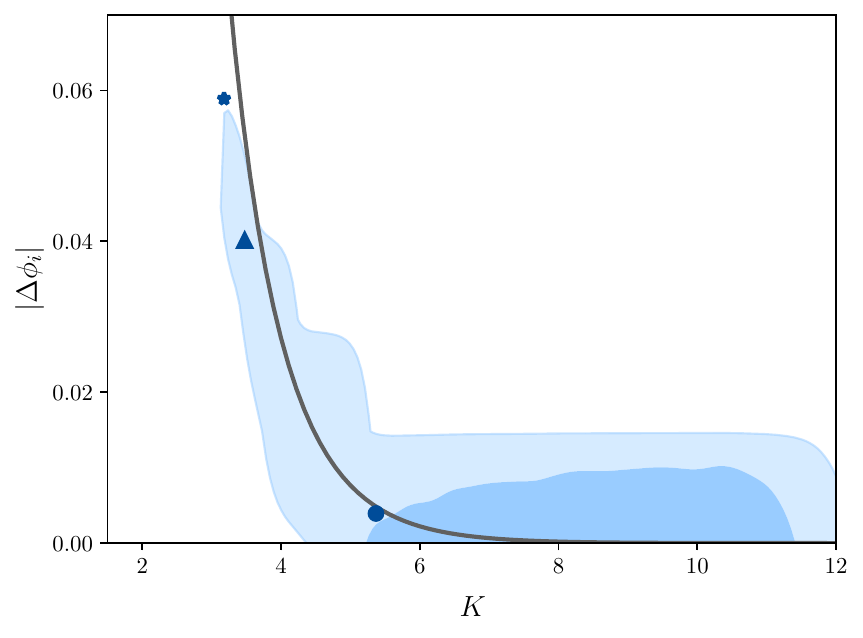}
    \caption{{Analytic results for $\Delta \phi_i$ obtained from eq.~\eqref{eq:phii}  for the hilltop quintessence models, and posterior contours furnished by a  MCMC analysis in the $K$-$\vert \Delta\phi_{i}\vert $ plane. 
    The analytic results are represented  in dark grey line using best-fit values for $\Omega _{\phi,0}$ and $w_0$ from the data combination with Union3.
    Dark blue shapes indicate the points corresponding to the best-fit values (see Tables \ref{tab:Axion_table_full}-\ref{tab:Higgs_table_full}) for model parameters ($\phi _0$ and $f$): circle for axion model, star for sugra model and triangle for the field theory model. 
    In the same figure, we show
    with blue contours 
    the $1\sigma$ and $2\sigma$ bounds in the $K-\vert \Delta\phi_{i}\vert $ plane from the constraints on the DS parameterisation for the data combination with Union3.  See Figure \ref{fig:phiK_DS_all} in the appendix for the analogous figure including constraints from all the data combinations.}
    }
    \label{fig:phiK_DS}
\end{figure}

As we have anticipated, a problem particular to hilltop quintessence is the fine-tuning of initial conditions close to the dS maximum.  Intuitively, the higher the curvature of the hilltop, the closer to the hilltop $\phi$ needs to start
in order to drive cosmic acceleration.  Even if such initial conditions can be selected via some physical mechanism in the early universe before big bang nucleosynthesis  -- e.g.~dynamically or anthropically -- they need to subsequently survive quantum diffusion effects. 
 The DS parameterisation provides us with an analytical expression for the initial displacement from the top of the potential,
$\Delta\phi_i$, in terms of the curvature of the potential at the hilltop, $K$,  the equation of state  parameter today, $w_0$, and the density parameter for quintessence today, $\Omega_{\phi,0}$:

\bea\label{eq:phii}
\Delta\phi_i &=& 4K\Omega_{\phi,0}\sqrt{\frac{(1+w_0)}{3}}\frac{(1-\Omega_{\phi,0})^{\frac{K-1}{2}}}{\left(K \sqrt{\Omega_{\phi,0}}-1\right)\left(1+\sqrt{\Omega_{\phi,0}}\right)^K +\left(K \sqrt{\Omega_{\phi,0}} + 1\right)\left(1-\sqrt{\Omega_{\phi,0}}\right)^K}\,. \nonumber \\ 
\eea
Note that as $w_0$ increases from $-1$ to $1$, $\Delta\phi_i$ also grows, and as $\Omega_{\phi,0}$ increases from 0 to 1, $\Delta\phi_i$ decreases. Although the overall dependence on $K$ is complex, it is clear that  the larger the value of $K$ (function of the hilltop curvature in \eqref{eq:K}), the larger the curvature, and therefore, the smaller initial displacement from the hilltop (smaller $\vert \Delta\phi_{i}\vert $) is expected.  
We make  manifest such behaviour with Figure~\ref{fig:phiK_DS}, where we plot the $K$-dependence of $\vert \Delta\phi_{i}\vert $ 
 using the best-fit values for $w_0$ and $\Omega_{\phi_0}$ (see Table \ref{tab:DS_table_full}) from the cosmological analysis described in Subsection \ref{MCMCDS}, together with the derived $1\sigma$ and $2\sigma$ limits on  the derived parameter $|\Delta\phi_i|$.
In the same plot we include the predictions for different hilltop quintessence models.
 
The $2\sigma$ tail extending towards larger $|\Delta\phi_i|$ in Figure~\ref{fig:phiK_DS} is expected, since very small curvature is still allowed when $\phi _i$ is far away from the hilltop, and, in fact, this tail follows the analytical curve. This asymmetric part in $2\sigma$ eventually leads to the skewed posterior distribution for small $K$ values shown in Figure~\ref{fig:DS0} in next section.

Once the 
 best-fit  value of the displacement from the hilltop inferred from the cosmological data has been found,  this in turn sets an upper bound on the Hubble scale at the end of reheating, if we assume that the initial conditions were set up by this time.  Indeed, for quantum diffusion at around the time of reheating -- and thus any time thereafter -- not to kick $\phi$ too far away from the value inferred from observations, we require: 
\be
H_{\rm rh} \ll 2 \pi \Delta \phi_{i
} 
\,.
\ee
For example, choosing the mean values for the DS
parameterisation in Table \ref{tab:param_limits_DS0} for the data combination with Union3, 
we obtain an upper bound on the reheating scale, $H_{\rm rh} \ll 0.06 M_{\rm Pl}$. We will obtain analog  bounds in our model-by-model analyses below.

\section{Cosmological analysis}
\label{sec_cosan}

We modify the cosmological Boltzmann code \texttt{CAMB} to implement the three hilltop models as well as the DS parameterisation described above. For each model, we perform a Markov Chain Monte-Carlo (MCMC) analysis of the parameter space, varying two model specific parameters (described below) alongside the baseline cosmological parameters \mbox{$\{\Omega_{\rm b}h^2,\Omega_{\rm c}h^2,H_0,\tau,A_s,n_s\}$} for which we adopt wide uniform priors. We make use of the following datasets:
 \begin{enumerate}
    \item{CMB from \textit{Planck} 2018:}
\begin{itemize}
    \item[-] \textit{Planck} 2018 low-$\ell$ temperature and polarisation likelihood~\cite{Aghanim:2019ame}.
    \item[-]  \textit{Planck} high-$\ell$ CamSpec TTTEEE temperature and polarization likelihood using \texttt{NPIPE} (\textit{Planck} PR4) data~\cite{Rosenberg:2022sdy}.
    \item[-] \textit{Planck} 2018 lensing likelihood~\cite{Aghanim:2018oex}.
\end{itemize}
 Hereafter, we collectively refer to all the \textit{Planck} CMB likelihoods as `CMB'.
    \item BAO likelihoods from DESI DR1~\cite{DESI:2024lzq,DESI:2024mwx,DESI:2024uvr} 
     \item  Pantheon+~\cite{Brout:2022vxf}, Union3~\cite{Rubin:2023ovl} and DES-Y5~\cite{DES:2024tys} type Ia supernovae likelihoods. 
    
\end{enumerate}
We sample the likelihoods using the \texttt{MCMC} sampler~\cite{Lewis:2002ah,Lewis:2013hha}, provided in \texttt{Cobaya}~\cite{Torrado:2020dgo}. Our convergence criteria for the MCMC chains is reached at the value $R-1=0.02$ for the Gelman-Rubin diagnostic. The constraints and posterior distribution plots for each model are generated using the~\texttt{GetDist} package~\cite{Lewis:2019xzd}. We also run the \texttt{Py-BOBYQA}~\cite{Bobyqa1,Bobyqa2} minimizer via \texttt{Cobaya} to obtain the maximum likelihood point and the corresponding $\chi^2$ values.

\smallskip

Recent results from the DESI BAO analysis~\cite{DESI:2024mwx}, alone as well as when combined with supernovae data from Pantheon+~\cite{Brout:2022vxf}, Union3~\cite{Rubin:2023ovl} and DESY5~\cite{DES:2024tys}, exhibit a preference for {dynamical} dark energy with a fairly rapid evolution in the recent past 
\cite{DESI:2024mwx,Calderon:2024uwn,Lodha:2024upq,RoyChoudhury:2024wri}.     
The significance of this deviation from $\Lambda$CDM ranges from $2-4\sigma$, depending upon the supernovae dataset chosen.\footnote{Different interpretations of these results as well as  their various cosmological implications are also discussed in~\cite{Colgain:2024xqj,Carloni:2024zpl,Park:2024jns,Wang:2024rjd,Cortes:2024lgw,Wang:2024pui,Dinda:2024kjf,Croker:2024jfg,Wang:2024hks,Luongo:2024fww,Mukherjee:2024ryz,Wang:2024dka,Efstathiou:2024xcq,Tada:2024znt,Yin:2024hba,Berghaus:2024kra,Shlivko:2024llw,Alestas:2024eic,Sohail:2024oki,Andriot:2024sif}.} When it comes to the quintessence models considered here, these datasets allow us to  provide constraints on the underlying model parameters as well as test whether these models can provide a better fit to the data compared to $\Lambda$CDM, or to  the CPL parameterisation~\citep{Giare:2024ocw,Giare:2024gpk,Wolf:2023uno}.

\subsection{Axion hilltop}
\label{sec_axhilc}

\begin{figure}[H]
    \centering
    \includegraphics[width=0.75\linewidth]{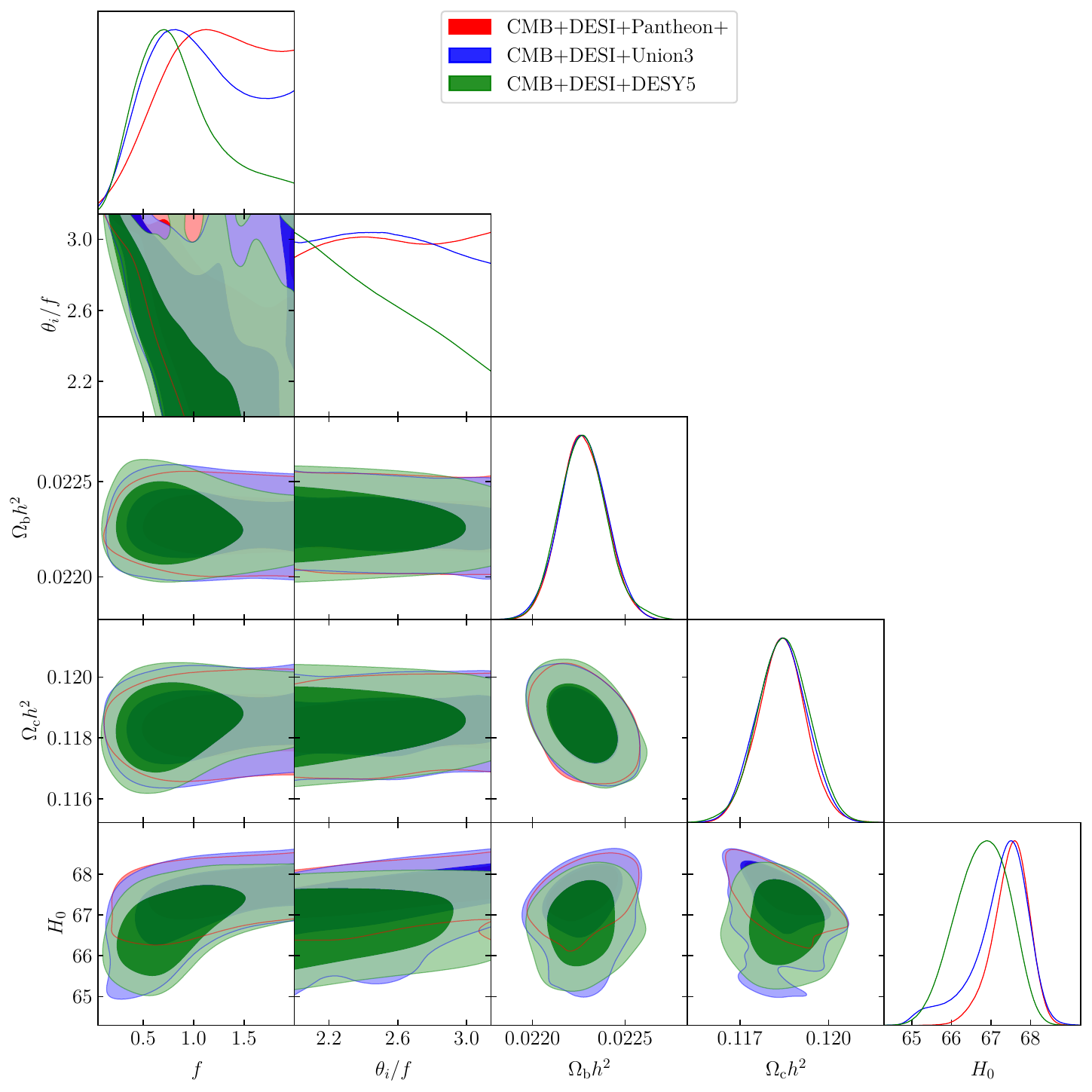}
    \caption{Parameter constraints on the Axion model, eq.~\eqref{axionV} ($68\%$ and $95\%$ contours).}
    \label{fig:cos_mcmc0}
\end{figure}

For the axion model discussed
in Section \ref{sub_string} we sample the axion decay constant $f$ and the initial field value
\footnote{The amplitude of the potential $V_0$ is tuned by the code and adjusted to obtain the correct $\Omega_\phi$ today.}
$\theta_i$ rescaled by $f$, i.e. $\theta_i/f$, for $f <2$. The results are plotted in Figure \ref{fig:cos_mcmc0} and $68\%$  limits summarised in Table \ref{tab:param_limits_cos}  
for the parameters\footnote{For each model studied in this section, the results  for the full set of parameters including the best-fit parameter combinations are presented in Appendix~\ref{app:fullmcmc}. In this section, we focus mainly on the dark energy model specific and the cosmological background parameters as, in any case, the constraints on the other cosmological parameters do not differ significantly across the different models or the different datasets.}  $\{f,\theta_i/f,\Omega_{\rm b}h^2,\Omega_{\rm c}h^2,H_0\}$.  
Focusing on the 1D marginalised constraints for the parameter $f$,   
we notice a preference for larger $(f\gtrsim 1)$ values in the Pantheon+ dataset, which decreases progressively as we change the supernovae dataset to Union3 or DESY5.  
In addition, from the constraints in the ($f$-$\theta_i/f$) plane  using the DESY5 SN dataset, we learn that for smaller $f$, the allowed values of $\theta_i/f$ are squeezed to a small region around $\pi$, while for larger $f$ the region around $\theta_i/f=\pi$ is excluded. This happens because for larger $f$ (smaller slope), one has to start farther away from the hilltop (maxima) to obtain dynamical dark energy at the present epoch. These effects are much less pronounced for the other SN datasets, reflecting the fact that these do not deviate from $\Lambda$CDM as much as DESY5. In other words, for the DESY5 dataset, the preferred field evolution in the axion model requires either the field starting far away from the maximum, if $f$ is large, or the field starting close to the maximum, if $f$ is small.

Table~\ref{tab:param_limits_cos} indicates that  the combined data sets give a lower bound on $f$ at around $f \gtrsim 0.7$  ($68\%$ C.L), and including the DESY5 data gives a mean value $f=0.88^{+0.24}_{-0.54}$. Note that the lower-limits  
and means presented here derive from a Bayesian analysis of the model against the cosmological data, thus they are highly prior-dependent. As we can see in both Figure \ref{fig:cos_mcmc0} and Table~\ref{tab:param_limits_cos}, the data are  not particularly constraining when it comes to the parameters $f,\,\theta_i/f$ and -- as long as this is the case -- the prior dependence of the limits is expected to remain. On the other hand, refinement of the theory priors on $f$ (or even $\theta_i$) will lead to tighter constraints on these parameters.  
 If evidence for dynamical dark energy persists, the parameter region $f>1$ with $\theta_i$ close to the maximum will be strongly disfavoured.

Motivated by the swampland constraints discussed in Section \ref{sec_theory}, which suggest 
{$f \lesssim \mathcal{O}(1) \Mp$}, we have assumed the prior that $f<2 \Mp$, restoring Planck units.  It would be important to refine the order one constants that appear in the swampland constraints.  Nevertheless,  
 the values of $f$ favoured by  
 our analysis 
  are rather large from the string theory point of view {and they could only be pushed further up by extending the priors to allow larger values for $f$, until the data is sufficiently constraining to make the fits prior-independent}.   In particular, for such large values of\footnote{Large values of $f$ also imply large values for $\theta_i$ in Planck units.  The distance conjecture \cite{Ooguri:2006in} generally puts super-Planckian field ranges in the swampland, but, in our hilltop scenario, the field range actually explored would be small due to Hubble friction.} $f \gtrsim 0.7 \Mp$, the weak gravity conjecture \eqref{wgc} implies that the instanton that generates the scalar potential has action $S_{\rm inst} \lesssim 1.4$.  This means that an additional source of exponential suppression is needed to achieve the hierarchically small scale of Dark Energy -- e.g. in the form of polyinstantons as discussed in \cite{Cicoli:2024yqh,Cicoli:2012tz}.  Recalling also that control of the instanton expansion requires $S_{\rm inst} \gtrsim 1$, theoretical and observational constraints combine to give a very narrow window of possibilities that would have to rely on numerical control\footnote{We finally note that for the best fit values we obtain for $f$ shown in Table \ref{tab:Axion_table_full}, we obtain a mass of the axion $m_\theta\approx 10^{-34}$eV. }.

Finally, it is interesting to consider the implications of the observational constraints on the initial conditions for the inflationary or reheating scale, by demanding that the initial conditions are safe from quantum diffusion at the time of reheating.  Following Section \ref{sec:bounds}, and using the mean values for $\theta_i$ and $f$ from the CMB+DESI+DESY5 data in Table \ref{tab:param_limits_cos} for illustration, we find, restoring $\Mp$:
\begin{equation}
    H_{\rm rh} \ll 2\pi(\pi f-\theta_i) = 4.1 \Mp\,,
\end{equation}
that is, there is no effective constraint on the reheating scale.

Note that the axion model has been analysed previously in \cite{Schoneberg:2023lun}, using the cosmological data available at that time. Our results with the new datasets are in agreement with their results, with the slight differences in the constraints on the axion parameters mainly driven by the new datasets.

\renewcommand{\arraystretch}{1.4}
\begin{table}[H]
    \centering
\begin{tabular} { |c| c| c| c|}

\hline
\rowcolor{gray!30} 
 \bf{Parameter} &  {\bf +Pantheon+} & {\bf +Union3} &  {\bf +DESY5} \\
\hline
{\boldmath$f              $} & $> 0.946                   $ & $> 0.779                   $ & $0.88^{+0.24}_{-0.54}      $\\

{\boldmath$\theta_i/f       $} & ---                          & ---                          & $< 2.62                    $\\

{\boldmath$\Omega_\mathrm{c} h^2$} & $0.11842\pm 0.00081        $ & $0.11842\pm 0.00083        $ & $0.11847\pm 0.00086        $\\

{\boldmath$\Omega_\mathrm{b} h^2$} & $0.02227\pm 0.00012        $ & $0.02227\pm 0.00013        $ & $0.02227\pm 0.00013        $\\

{\boldmath$H_0            $} & $67.49^{+0.51}_{-0.37}     $ & $67.23^{+0.81}_{-0.40}     $ & $66.79^{+0.74}_{-0.62}     $\\

$\theta_i                    $ & $3.1^{+1.1}_{-1.4}         $ & $2.73^{+0.93}_{-1.6}       $ & $2.11^{+0.40}_{-1.2}       $\\
\hline
\end{tabular}
    \caption{Axion model: parameter means and $68\%$ limits for the addition of the different supernovae datasets to the CMB+DESI combination.}
    \label{tab:param_limits_cos}
\end{table}

\subsection{Saxion hilltop}

\begin{figure}[H]
    \centering
    \includegraphics[width=0.75\linewidth]{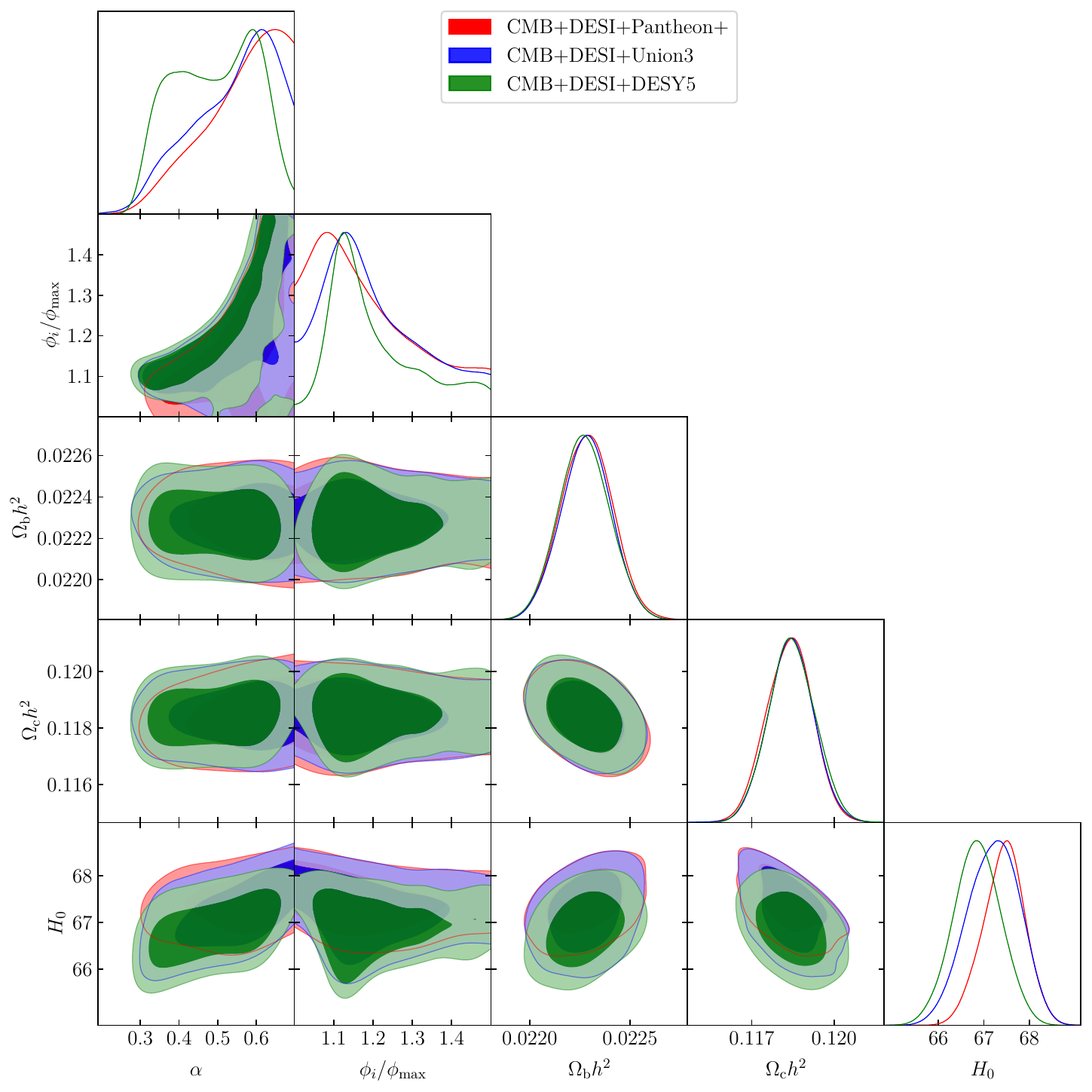}
    \caption{Parameter constraints ($68\%$ and $95\%$ contours) for the Saxion model \eqref{sugraV}. }
    \label{fig:sugra_mcmc0}
\end{figure}

For the saxion model we sample the parameter $\alpha\in \lp \frac{2\pi}{32}, \frac{2\pi}{9}\rp$, consistent with our theory discussion in Section \ref{sec_theory}, and the initial field value 
$\phi_i$, rescaled by $\phi_{\rm max}$, i.e.~$\phi_i/\phi_{\rm max}$. 
The results are plotted in Figure \ref{fig:sugra_mcmc0} and the $68\%$  limits summarised in Table \ref{tab:param_limits_sugra} 
for the parameters $\{\alpha,\phi_i/\phi_{\rm max},\Omega_{\rm b}h^2,\Omega_{\rm c}h^2,H_0\}$. 

We interpret these parameter constraints in terms of the series expansion of the Saxion potential in~\eqref{eq:Vsugraexp}, where we learn that $\alpha$ effectively plays the role of the  potential's curvature  through the term $\phi_{\rm max}$. Since the field value at the potential maximum is a function of $\alpha$ \eqref{eq:fimax}, the potential has a  higher curvature for smaller $\alpha$ and  lower curvature for larger $\alpha$. Thus, for smaller $\alpha$ the parameter $\phi_i/\phi_{\rm max}$ is constrained to be close to one. This manifests as a somewhat curved degeneracy in the $\alpha$--$\phi_i/\phi_{\rm max}$ plane in Figure \ref{fig:sugra_mcmc0}. Focusing in particular on the DESY5 contours (green), whilst $\phi_i/\phi_{\rm max}$ moves towards smaller values as $\alpha$ decreases, it cannot be too close to 1 as otherwise one would not obtain dynamical dark energy at the present epoch, as is preferred by the DESY5 data.  
Although current data do not put any tight constraint on the fundamental parameter $\alpha$, if e.g.~$\phi_i$ is forced to be closer to the hilltop and upcoming surveys continue to see evidence for dynamical dark energy, the preferred region for  $\alpha$  would be driven towards lower values.

Finally, we note the constraints on the reheating scale, by demanding that the initial conditions are safe from quantum diffusion at that time.  Following Section \ref{sec:bounds}, and using the mean values for $\phi_i$ and $\alpha$ from the CMB+DESI+DESY5 data in Table \ref{tab:param_limits_sugra} for illustration, we find, restoring $\Mp$:
\begin{equation}
    H_{\rm rh} \ll 2\pi\left(\phi_i-\frac{1}{\sqrt{2}}\log\left(\frac{1}{\sqrt{2}\alpha}\right)\right) = 0.38 \Mp\,.
\end{equation}

\renewcommand{\arraystretch}{1.4}
\begin{table}[H]
    \centering
\begin{tabular} { |c| c| c| c|}

\hline
\rowcolor{gray!30} 
 \bf{Parameter} &  {\bf +Pantheon+} & {\bf +Union3} &  {\bf +DESY5} \\
\hline
{\boldmath$\alpha         $} & $> 0.521                   $ & $0.537^{+0.16}_{-0.045}    $ & $0.49\pm 0.11              $\\

{\boldmath$\phi_{i}/\phi_{\mathrm{max}}$} & $< 1.23                    $ & $1.201^{+0.071}_{-0.17}    $ & $1.205^{+0.058}_{-0.14}    $\\

{\boldmath$\Omega_\mathrm{c} h^2$} & $0.11838\pm 0.00084        $ & $0.11842\pm 0.00083        $ & $0.11846\pm 0.00085        $\\

{\boldmath$\Omega_\mathrm{b} h^2$} & $0.02228\pm 0.00013        $ & $0.02228\pm 0.00013        $ & $0.02228\pm 0.00012        $\\

{\boldmath$H_0            $} & $67.44^{+0.49}_{-0.43}     $ & $67.21^{+0.62}_{-0.55}     $ & $66.86\pm 0.52             $\\

$\phi_i                    $ & $0.209^{+0.064}_{-0.18}    $ & $0.241^{+0.083}_{-0.21}    $ & $0.32^{+0.13}_{-0.22}      $\\
\hline
\end{tabular}
    \caption{Saxion model: parameter means and $68\%$ limits for the addition of the different supernovae datasets to the CMB+DESI combination. The full set of constraints can be found in Table~\ref{tab:Sugra_table_full} and corresponding plots in Figure~\ref{fig:sugra_mcmc}.}
    \label{tab:param_limits_sugra}
\end{table}

\subsection{Higgs-like hilltop}

\begin{figure}[H]
    \centering
    \includegraphics[width=0.75\linewidth]{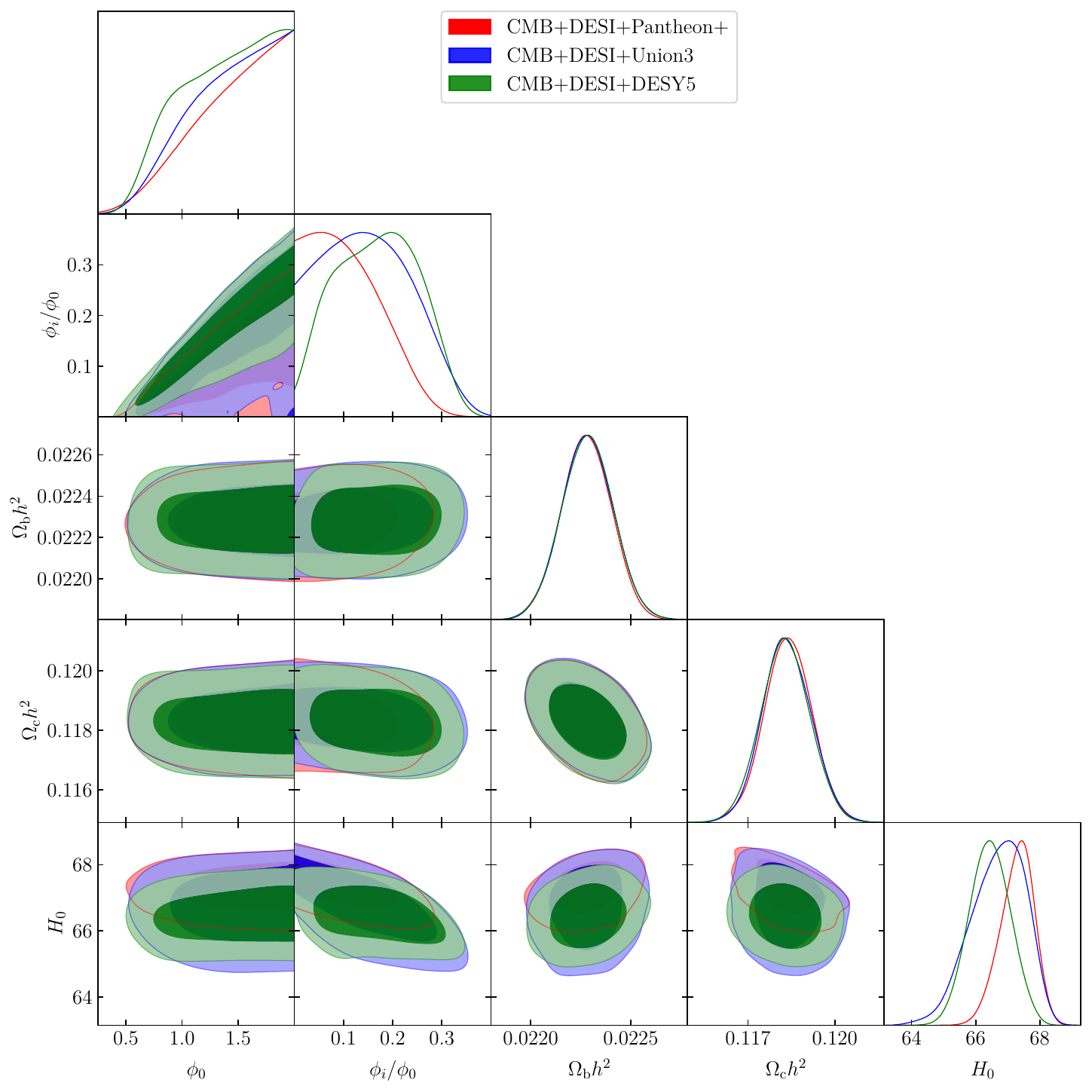}
    \caption{Constraints on the Higgs-like hilltop model \eqref{eq:FTV} ($68\%$ and $95\%$ contours). }
    \label{fig:Higgs_mcmc0}
\end{figure}

For the Higgs-like hilltop\footnote{See \cite{Wolf:2024eph} for a cosmological analysis of a pure quadratic hilltop.}, we vary the parameter $\phi_0$ and the initial relative field value $\phi_i/\phi_0$. In fact these two quantities play a role similar to $f$, $\theta_i/f$ in the Axion model, representing the steepness of the potential and the initial displacement from the maximum. The resulting parameter posterior distributions can also be interpreted similarly and are plotted in Figure \ref{fig:Higgs_mcmc0}, with the $68\%$ limits summarised in Table \ref{tab:param_limits_Higgs}. In particular, we notice the same squeezing of the allowed $\phi_i$ when the potential is steep, i.e.~for smaller $\phi_0$, as found in the 
axion model of Section \ref{sec_axhilc}. 
 The limits from the current cosmological data are strongly prior dependent and once again, a more precise understanding of the theoretical priors on $\phi_0$ and $\phi_i$ will be crucial to constraining the parameter space of this model.

Finally, we estimate the constraints on the reheating scale from quantum diffusion and the degree of fine-tuning in the initial conditions that the data indicates.  Following Section \ref{sec:bounds}, and using the mean value for $\phi_i$ from the CMB+DESI+DESY5 data in Table \ref{tab:param_limits_Higgs} for illustration, we find, restoring $\Mp$:
\begin{equation}
    H_{\rm rh} \ll 2\pi\phi_i = 1.6 \Mp\,,
\end{equation}
that is, no effective constraint on the reheating scale.

\renewcommand{\arraystretch}{1.45}
\begin{table}[H]
    \centering
\begin{tabular} { |c| c| c| c|}

\hline
\rowcolor{gray!30} 
 \bf{Parameter} &  {\bf +Pantheon+} & {\bf +Union3} &  {\bf +DESY5} \\
\hline
{\boldmath$\phi_0         $} & $> 1.29                    $ & $> 1.24                    $ & $> 1.17                    $\\

{\boldmath$\phi_i/\phi_0  $} & $< 0.142                   $ & $0.151^{+0.073}_{-0.12}    $ & $0.169\pm 0.081            $\\

{\boldmath$\Omega_\mathrm{c} h^2$} & $0.11838\pm 0.00082        $ & $0.11835\pm 0.00084        $ & $0.11829\pm 0.00084        $\\

{\boldmath$\Omega_\mathrm{b} h^2$} & $0.02228\pm 0.00012        $ & $0.02228\pm 0.00013        $ & $0.02228\pm 0.00013        $\\

{\boldmath$H_0            $} & $67.29^{+0.59}_{-0.45}     $ & $66.7^{+1.0}_{-0.70}       $ & $66.44\pm 0.64             $\\

$\phi_i                    $ & $0.174^{+0.071}_{-0.17}    $ & $0.235^{+0.088}_{-0.23}    $ & $0.26^{+0.10}_{-0.24}      $\\
\hline
\end{tabular}
    \caption{Higgs-like hilltop model: parameter means and $68\%$ limits for the addition of the different supernovae datasets to the CMB+DESI combination. }
    \label{tab:param_limits_Higgs}
\end{table}

\subsection{DS parameterisation analysis}
\label{MCMCDS}

\begin{figure}[H]
    \centering
    \includegraphics[width=0.85\linewidth]{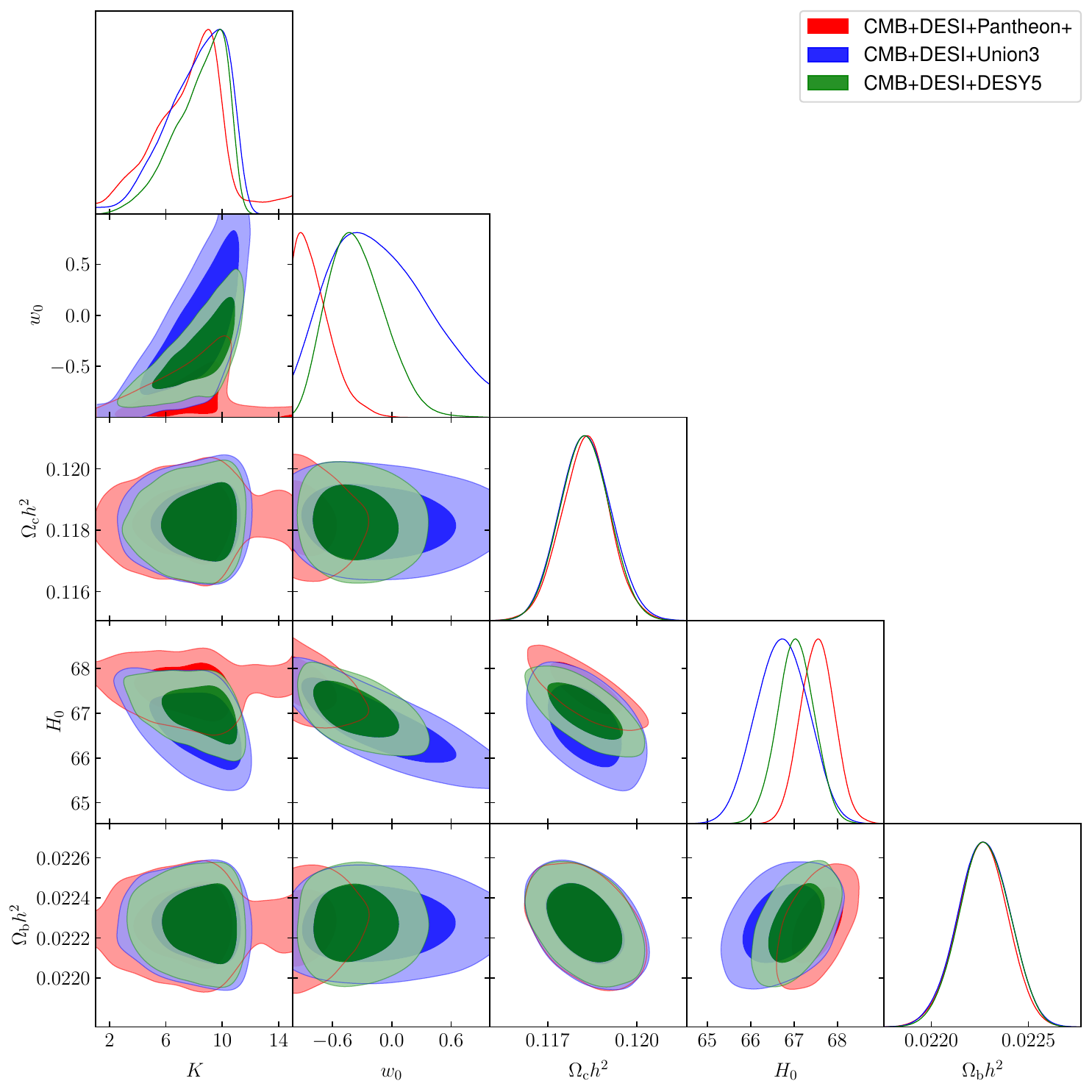}
    \caption{DS parameterisation: parameter means and  limits for the addition of the different supernovae datasets to the CMB+DESI combination.  }
    \label{fig:DS0}
\end{figure}

For the DS parameterisation, we vary the parameters $K$ and $w_0$ along with the other six CDM parameters as in the previous sections. The results for the posterior probability distributions are plotted in Figure~\ref{fig:DS0} and the corresponding $68\%$ limits are presented in the Table~\ref{tab:param_limits_DS0}. The full set of posterior distributions and corresponding limits are reported in Figure~\ref{fig:DSfull} and Table~\ref{tab:DS_table_full}, respectively. The 1D posterior distributions of $K$ for all data combinations are skewed towards smaller values -- as   expected since smaller curvature  
 is  required when the initial field value is far from the hilltop (see also the discussion around Figure~\ref{fig:phiK_DS}). For what concerns the combination with Pantheon+ results, 
the large $K$ tail of the 1D posterior  is  a consequence of the condition  $w_0\simeq -1.0$ being 
preferred when this data set is added\footnote{Our prior range for $w_0$ excludes the $w_0< -1.0$ region.}.

Although a one-to-one match among the posteriors of $f$ and $\phi_0$ with that of $K$ is not expected due to the non-linear relation between these parameters~\eqref{eq:Kax}, and~\eqref{eq:Kft} -- as well as the complicated mapping between $\phi_i$ and $w_0$ --   the values of $K$ obtained directly from $f$ and $\phi _0$ can nevertheless be compared with the posterior of $K$ from DS parameterisation. Using the mean values for $f$ and $\phi _0$  from Tables~\ref{tab:param_limits_cos} and~\ref{tab:param_limits_Higgs} (for +Union3) leads to $K_{\rm ax}^{\rm mean} < 1.45$, and $K_{\rm Higgs}^{\rm mean}<2.11$, while $K_{\rm sugra}=3.179$ is independent of model parameters \footnote{ Using best-fit values for $f$ and $\phi _0$ from Tables~\ref{tab:Axion_table_full} and~\ref{tab:Higgs_table_full} respectively (for +Union3), viz, $K_{\rm ax}^{\rm best-fit} = 5.3685$, and $K_{\rm Higgs}^{\rm best-fit}=3.4765$.}. However, $K$ in DS parameterisation is not very tightly constrained for any data combination, as compared to its prior range, allowing for the values of $K$ from the individual model constraints to lie within $1\sigma$ -- $3\sigma$ of their DS counterparts,  depending on the data combination under consideration. This can {also be} clearly seen in Figure~\ref{fig:phiK_DS} in the relative position of the dots (for model predictions) with respect to the shaded 1$\sigma$ and 2$\sigma$ regions in the $K$-$|\Delta\phi_i|$ plane.

A similar analysis for the DS parameterisation can be found  in~\cite{Gialamas:2024lyw}, where a number of parameters including $K$ have been varied, while CDM parameters were kept fixed at CMB values.  This analysis obtained even larger values for $K$, albeit with large error.
 Our results and those in~\cite{Gialamas:2024lyw} are consistent within $2\sigma$ error. 

Finally, we note that $w_0$ itself is also poorly constrained. Due to the  weak constraints on the DS parameters $w_0,\,K$, the derived constraints on $\vert \Delta \phi _i\vert$ shown in Figure~\ref{fig:phiK_DS} also allow for a large range of values.  Since $w_{\phi}$ is best constrained at around $z \approx 0.4$ -- as manifest also from the DESI reconstruction plotted in Figure~\ref{fig:bestfit_v_reconstruction} below -- we also present the derived constraints on $w_{\phi}(z=0.4)$ in Figure~\ref{fig:wzp4_double} in Appendix \ref{app:fullmcmc} as 2D posterior plots from the analysis of DS parameterisation. In particular, the $w_0$-$w_{\phi}(z=0.4)$ plot demonstrates that the constraints on $w_{\phi}(z=0.4)$ are improved by at least an order of magnitude as compared to the constraints on $w_0$ only.

\renewcommand{\arraystretch}{1.45}
\begin{table}[H]
    \centering
\begin{tabular} { |c| c| c| c|}

\hline
\rowcolor{gray!30} 
 \bf{Parameter} &  {\bf +Pantheon+} & {\bf +Union3} &  {\bf +DESY5} \\
\hline
{\boldmath$K$} & $ 7.6^{+2.5}_{-2.1}             $ & $8.2^{+2.7}_{-1.4}    $ &  $8.4^{+2.4}_{-1.1}$\\

{\boldmath$w_0$} &   $<-0.709$        &   $-0.11^{+0.39}_{-0.61}$      & $-0.35^{+0.24}_{-0.33}$     \\

{\boldmath$\Omega_\mathrm{c} h^2$} & $0.11830\pm 0.00081         $ & $0.11828 \pm 0.00086      $ & $0.11825\pm 0.00082$  \\

{\boldmath$\Omega_\mathrm{b} h^2$} & $0.02226\pm 0.00013$ & $0.02227\pm 0.00013$ & $ 0.02227\pm 0.00013$\\
{\boldmath$H_0            $} & $ 67.53\pm0.39 $ & $66.73\pm 0.59     $ & $67.04\pm 0.42 $\\
\hline
\end{tabular}
    \caption{DS parameterisation: parameter means and $68\%$ limits for the addition of the different supernovae datasets to the CMB+DESI combination. }
    \label{tab:param_limits_DS0}
\end{table}

\subsection{Model comparison}\label{sec:modelcomp}
To compare the quality of the fit to the data provided by the different models, we perform a  model comparison based
on the Akaike information criterion (AIC)~\cite{AIC,Liddle:2004nh}, which also takes into account the number of free parameters of the model. The AIC value is defined as follows
\begin{align}
    \mathrm {AIC} \,=\,2n-2\ln \mathcal{L}_{\rm max}\,,
\end{align}
where $\mathcal{L}_{\rm max}$ stands for the maximum likelihood value in the model and $n$ the number of free parameters. Models with smaller AIC are favoured by the data with the best model having the lowest AIC value. To compare between models, one looks at the difference in AIC values with $\Delta \mathrm{AIC}_{12} {\equiv \mathrm{AIC}_1 - \mathrm{AIC}_2} \lesssim 2$ indicating no preference between Model 1 and Model 2, whilst $\Delta \mathrm{AIC}_{12} \gtrsim 5$ indicates a strong preference for Model 2 over Model 1.

\begin{figure}[H]
    \centering
    \includegraphics[width=0.49\linewidth]{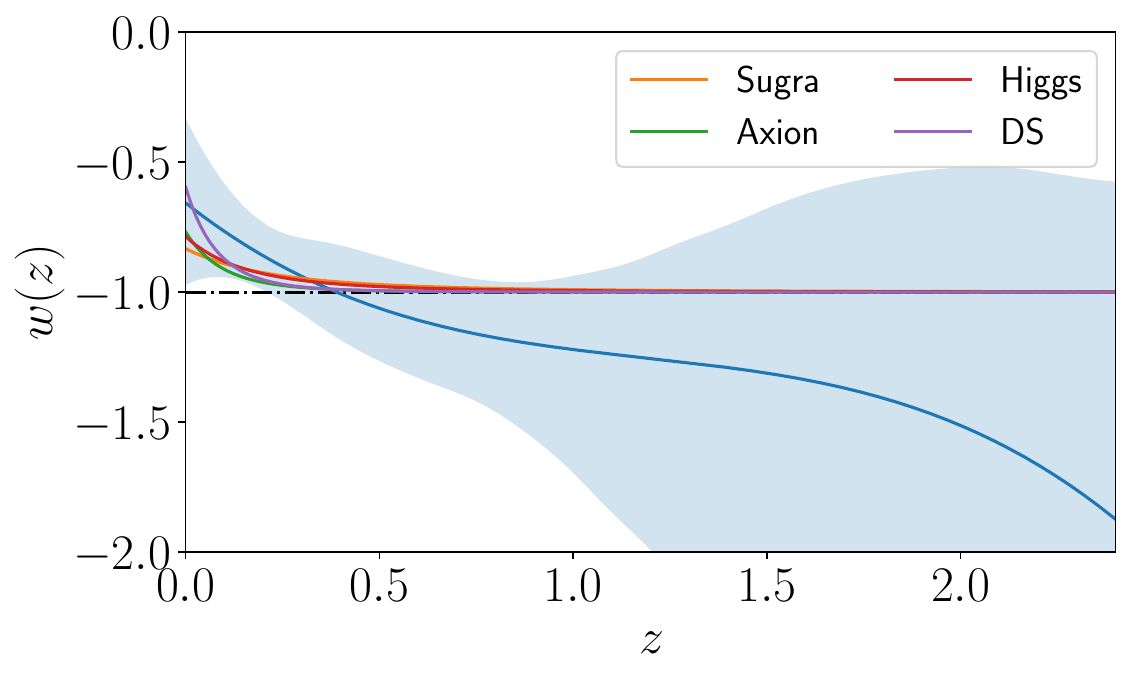}
        \includegraphics[width=0.49\linewidth]{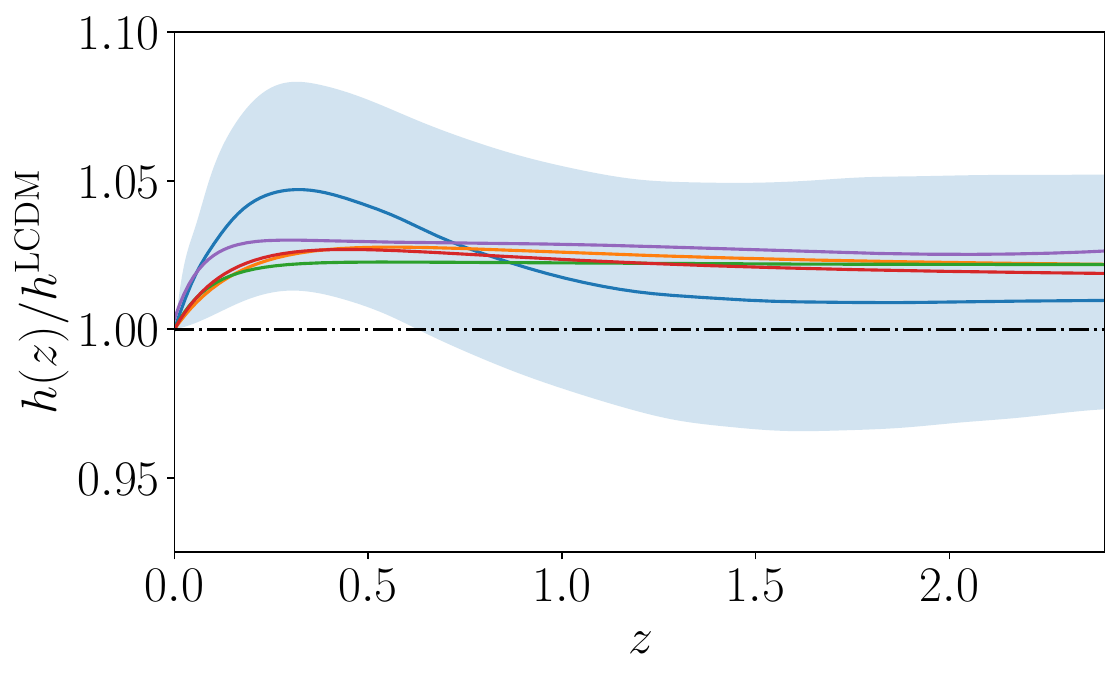}
    \caption{The quantities  $w_{\rm \phi}(z)$, and $h(z)\equiv H(z)/H_0$ are plotted for the best-fit   hilltop models as indicated, and compared to the DESI reconstruction (blue line) using CMB+DESI+Union3 data~\cite{Calderon:2024uwn}. {The CPL parameterisation matches the DESI reconstruction.}  The shaded regions represent the 95\% confidence regions. At $z=0$, $w_{0} = -0.65$ for the reconstruction, while the values for the hilltop models are as follows $w_0=-0.77$ for the axion model, $w_0=-0.83$, for the saxion and $w_0=-0.79$ for the Higgs-like hilltop. For comparison, $w_{0}^{\exp} = -0.89$ for the exponential model (see Figure 10 of \cite{Bhattacharya:2024hep}). For the DS parameterisation, the best-fit value is $w_0^{\rm DS}=-0.60$. The best-fit parameter values for each model can be found in Appendix \ref{app:fullmcmc}.}
    \label{fig:bestfit_v_reconstruction}
\end{figure}

The AIC values for the different models we have studied are provided in Table~\ref{tab:model_comp}{, where we also consider $\Lambda$CDM, the CPL parameterisation and the exponential runaway quintessence model}. We learn that the preference for the hilltop quintessence models over $\Lambda$CDM is strongest for the dataset combination CMB+DESI+DESY5 and weakest for CMB+DESI+Pantheon+. Out of all the different models studied here, the CPL parameterisation remains the most favoured, irrespective of the dataset combination chosen. However, its improvement  with respect to the DS parameterisation is data dependent, with best improvement for +Union3 ($\Delta $AIC$_{\rm DS,CPL}\simeq 8.3$) and only  mild improvement for +DESY5 ($\Delta $AIC$_{\rm DS,CPL}\simeq 3.4$). We  attribute this  improvement to the  more rapid evolution of the dark energy equation of state in this model, as well as the phantom-like behaviour in the past~\cite{DESI:2024mwx}, which matches very well the DESI reconstruction of background quantities $w(z)$ and \mbox{$h(z)\equiv H(z)/H_0$}, shown in Figure~\ref{fig:bestfit_v_reconstruction}.  As is clear from Figure~\ref{fig:bestfit_v_reconstruction}, these two features cannot be produced in the hilltop quintessence models.  On the other hand, the redshift evolution seen here for the hilltop quintessence models is much closer to the DESI reconstruction than what can be produced in the exponential potential model  (compare  Figure~10 of \cite{Bhattacharya:2024hep}). At the same time,  Table~\ref{tab:model_comp} tells us that the overall improvement in the fit compared to the exponential model is not significant enough to compensate for the additional parameter introduced by the hilltop models. 

We  also notice that  the DS parameterisation fares (marginally) better than the concrete hilltop quintessence models we considered, with the  greatest improvement for +DESY5 with $\Delta $AIC$_{\rm model,DS}\simeq 4$, and the  lowest improvement for +Pantheon+ with $\Delta $AIC$_{\rm model,DS}\lesssim 0.6$. 
One can then 
  also adopt the approach to 
start from the DS parameterisation, 
 compare it with data, and infer 
the preferred values for the  curvature parameter $K$. We can then use these results 
to deduce a preference
for specific hilltop scenarios, given 
that each model 
 favours its own preferred range of $K$, depending 
 on each model parameters. The evolution of the background parameters plotted in Figure~\ref{fig:bestfit_v_reconstruction} shows clearly that the DS parameterisation tracks the predictions from hilltop quintessence models well until very low redshift, $z\gtrsim 0.1$, beyond which it leads to a faster evolution of $w(z)$.
 
\begin{table}[H]
    \centering
\resizebox{\textwidth}{!}{\begin{tabular}{|c|c|c|c|c|c|c|c|}
\hline
\rowcolor{gray!30}    \textbf{AIC} & Axion & Sugra & Higgs  & DS &  $\Lambda$CDM & CPL & Exp \\
    \hline 
     CMB+DESI+Pantheon+ & 12409.55 &  12409.40 & 12409.07  & 12408.9  & 12406.04 & 12401.70 & 12407.19\\
     \hline 
     CMB+DESI+Union3 & 11030.07 & 11029.49 & 11030.38  & 11027.9 & 11028.69 & 11019.62  & 11029.00\\
     \hline
     CMB+DESI+DESY5 & 12644.67 & 12645.65 &12644.89  & 12641.2 & 12649.01 & 12637.79 & 12644.73 \\
     \hline
\end{tabular}}
    \caption{Comparison between the different models considered in this section.}
    \label{tab:model_comp}
\end{table}

\section{Outlook and Future Challenges}
\label{sec_conc} 

Embedding models of dark energy in quantum gravity and string theory is a notoriously difficult task, given  stringent theoretical constraints on model building associated with quantum gravity conjectures, as well as strong phenomenological constraints on light fields. At the same time, recent and forthcoming cosmological results are going to probe the behaviour of dark energy  with an ever-increasing degree of precision,  offering the concrete  possibility to test our understanding of dark energy in quantum gravity with cosmological data.
 
 With the aim of exploiting such opportunities,
 in this work we considered a class of  dark energy models -- hilltop quintessence -- which is able to satisfy theoretical bounds from de Sitter and other quantum gravity conjectures. We examined various realisations of quintessence hilltops based on axions, their supersymmetric partners, and  Higgs-like string embeddings. Axion hilltops are widely considered and particularly well-motivated, as their shift symmetry evades problems with UV sensitivities, fifth-forces and time-variation of fundamental constants, and a priori -- at the hilltop -- they can be consistent with the weak gravity conjecture. Moreover, we have presented a dynamical mechanism by which hilltop initial conditions could be set up. However, it is important
 to point out that none of our scenarios solve the cosmological constant problem, hence from a theoretical viewpoint they are at least as fine-tuned as the $\Lambda$CDM model. 
  
 We studied  the cosmological consequences of our three string-motivated hilltop models and discussed a convenient parameterisation  of their associated equation of state. We then tested their predictions by means of a Bayesian  MCMC analysis   with 
   recent CMB, galaxy surveys, and supernova data. We  showed to what extent current data can distinguish amongst hilltop models and  impose constraints on their parameters. Interestingly, such experimental results are complementary to theoretical bounds from quantum gravity conjectures, and we discussed the consequences of these features for our current understanding of dark energy in string theory. Notably, observational constraints on the axion decay constant for axion hilltop quintessence are in tension with the weak gravity conjecture, illustrating how synergies between constraints from observations and from quantum gravity can rule out otherwise well-motivated models.  So far, model comparisons favour the CPL parameterisation over any of our hilltops, $\Lambda$CDM, and exponential runaways.\footnote{See e.g. \cite{Scherrer:2015tra} for possible maps of the CPL parameterisation to physical quintessence or barotropic dark energy models, though no quintessence potential will give rise to the phantom behaviour seen in Figure \ref{fig:bestfit_v_reconstruction}.} However CPL is only mildly improved with respect to the DS parametrisation, especially for the dataset CMB+DESI+DESY5.  At the same time, the limited constraining power of current data means that the model parameter constraints and comparisons that we obtained  are sensitive to our priors, which are based on theoretical assumptions about viable regions of parameter space. Consequently, given that our inference from cosmological  data   strongly
 depends  on theoretical assumptions, it is imperative to  refine our theoretical understanding of the priors, so as to  maximize the informative power of current and forthcoming cosmological datasets for testing dark energy scenarios in quantum gravity. In doing so, we can also hope that this endeavor will inspire new ideas and perspectives for tackling fundamental theoretical challenges, such as the cosmological constant problem in quantum gravity.

\smallskip
Assumptions entering into theoretical priors  constrain the size and location of the allowed region in  parameter space associated with a given model. This includes 
the possible values of parameters entering in  the model Lagrangian  and the allowed initial conditions or field ranges associated with  the dynamics of the 
quintessence scalar. Such information
can be theoretically refined by better specifying the `order one' constants\footnote{See e.g. \cite{Heidenreich:2015nta, Rudelius:2021oaz} for work towards fixing these order one constants using dimensional reduction.} entering the de Sitter
conjecture in  the inequalities \eqref{sdSC} and the weak gravity conjecture in \eqref{wgc}, or by embedding hilltop quintessence into more complete early universe models, able to accurately specify their  initial conditions. Also, the range of allowed  priors can be limited  by enriching the hilltop models to include additional Standard Model matter fields: then, one should take  into account further constraints on the parameter space from limits on  fifth forces and the time variation of fundamental constants. All these theoretical questions can be addressed by developments of the current theoretical tools at our disposal. We believe that such questions are very timely, and addressing them will allow  us to  exploit   the synergy between theory and observations offered by current and forthcoming cosmological probes. We look forward to continuing this analysis in forthcoming publications.

\section*{Acknowledgements}

The work of GB, AM, GT and IZ is partially funded by  STFC grant ST/X000648/1 and the work of SP is partially funded by STFC grant ST/X000699/1. SB acknowledges the ``Consolidaci\'on Investigadora'' grant CNS2022-135590 and her work is partially supported by the Spanish Research Agency (Agencia Estatal de Investigaci\'on) through the Grant IFT Centro de Excelencia Severo Ochoa No CEX2020-001007-S, funded by MCIN/AEI/10.13039/501100011033. We also acknowledge the support of the Supercomputing Wales project, which is part-funded by the
European Regional Development Fund (ERDF) via Welsh Government. For the purpose of
open access, the authors have applied a Creative Commons Attribution licence to any Author
Accepted Manuscript version arising. Research Data Access Statement: No new data were generated for this manuscript.

\newpage

\begin{appendix}
\section{Saxion-axion stringy hilltops}\label{app1}

In this appendix we summarise the  saxionic hilltop potential we consider in the main text \eqref{sugraV}. More details can be found in \cite{Olguin-Trejo:2018zun} (see also \cite{ValeixoBento:2020ujr}). We also provide a concrete example where the minima of stringy axions can become maxima upon turning on further subleading corrections, thus providing a possible dynamical mechanism that tunes the initial conditions of the axion to the hilltop.

\subsection{Saxion hilltops}
The model considers  some inflationary early Universe scenario that ends in a supersymmetric
Minkowski minimum, where most of the string moduli are stabilised and heavy, except for a flat direction corresponding to a chiral superfield $\Phi$. The K\"ahler potential for such flat direction takes the form:
\be
K=- n \ln{\left(\Phi+\bar\Phi\right)}\,,
\ee
where $n$ takes different values depending on the type of modulus. As we will see, we are interested in $n=1$, which may correspond to some (non-overall) volume modulus, or a complex structure modulus, or the dilaton. 
The superpotential is given by non-perturbative effects which lift the flat direction  at some scale before BBN. The leading term in the non-perturbative  superpotential is then given by
\be\label{eq:Wnp}
W_{\rm np}= A e^{-\alpha \,\Phi} \,.
\ee
Here $\alpha$ is a constant that can arise from different instanton types or gaugino condensation and can e.g.~take values $\alpha=2\pi/N$, with $N=1$ for an Euclidean D3-brane instanton  and $N  \geq 2$ for  gaugino  condensation with condensing group rank $SU(N)$. 
The scalar potential  can then be computed using the supergravity formula
\be\label{eq:Vsugra}
V= e^K\left[ K^{i\bar{j}}D_jWD_{\bar j}\bar W - 3|W|^2 \right]\,,
\ee
where $D_jW \equiv \partial_j W + K_j W$.  Writing the complex scalar field component of the chiral superfield as $\Phi=\varphi+i\theta$,  we obtain:
\be
V=\frac{A^2}{2^n n}\,e^{-2\,\alpha\,\varphi}\varphi^{-n}\left(n^2+n(-3+4\alpha \varphi) +4\alpha^2\varphi^2 \right)\,,
\ee
which, for $n=1$, has a dS maximum at:
\be\label{eq:varphimax}
\varphi_{\rm max} = \frac{1}{\sqrt{2} \alpha}\,.
\ee

Notice that at the leading order considered, the axion $\theta$ remains a flat direction, but it will be lifted by subleading non-perturbative terms \cite{Olguin-Trejo:2018zun}.  Indeed, adding a subleading contribution
\be\label{eq:Wsub}
W_{\rm np \, sub} = B e^{-\beta \Phi}\,
\ee
to the leading contribution \eqref{eq:Wnp}, will generate a minimum for the axion, whilst preserving the (slightly shifted) maximum in the saxion direction, with $|m_{\rm axion}^2|< |m_{\rm saxion}^2|$.  For example for $n=1$, $\beta = 2\alpha$, $B=-A/20$, the axion has a minimum at $\theta=\frac{2m\pi}{\beta-\alpha}$, $m\in {\mathbb Z}$ \cite{Olguin-Trejo:2018zun}.  Note that the exponential suppression of the subleading non-perturbative term is only by a factor $\sqrt{2}$ at the hilltop, so control of the expansion in non-perturbative effects can be at best numerical there.  
It is useful to express the potential in terms of the canonically normalised field, which for $n=1$, is 
\be
\phi = \Mp \sqrt{\frac{1}{2}}\log{\varphi}\,.
\ee
The final potential becomes \eqref{sugraV}.

\subsection{Subleading corrections to the axion}\label{app:suaxion}

Let us focus on the axion in the model considered above. As we saw, the axion is lifted after adding subleading correction \eqref{eq:Wsub}. Now further subleading corrections of the form
\be\label{eq:Wsubsub}
W_{\rm sub \, sub} =C e^{-\gamma \,\Phi}\,,
\ee
may turn the minimum of the axion potential at $C=0$ into a maximum when $C\ne 0$. This will happen for suitable values of the parameters $\alpha, \beta, \gamma$, and $A,B, C$. Schematically, the potential \eqref{eq:Vsugra} including the two subleading corrections, \eqref{eq:Wsub}, \eqref{eq:Wsubsub} 
takes the form 
\be\label{eq:subleadAxion}
V= g(\varphi_0) + f_1(\varphi_0) AB \,\cos\lp (\beta-\alpha)\theta\rp +
f_2(\varphi_0) AC \,\cos\lp (\gamma-\alpha)\theta\rp +
f_3(\varphi_0) BC \,\cos\lp (\gamma-\beta)\theta\rp \,,
\ee
where $g(\varphi_0), f_i(\varphi_0)$ are functions of the saxion (as well as $\alpha,\beta,\gamma$) evaluated at its extremum. 
For the case $C=0$ and $B<0$, as discussed above,  the potential has a minimum at $\theta_{\rm min}=\frac{2m\pi}{\beta-\alpha}$. When $C$ is turned on, this minimum can become a maximum for suitable values of the parameters. For example, for $n=1$ as above, $\alpha=2\pi/16$, $\beta=2\alpha$, $\gamma=3\alpha$, $B=-A/20$, $C=A/35$ ($A=1$ for concreteness), the minima for the axion at $C=0$ become  maxima for $C\ne0$.  
It is also possible that some minima stay minima, while only some become maxima. Of course, more complex modulated structures can arise. 
\begin{figure}[H]
    \centering
    \includegraphics[width=0.55\linewidth]{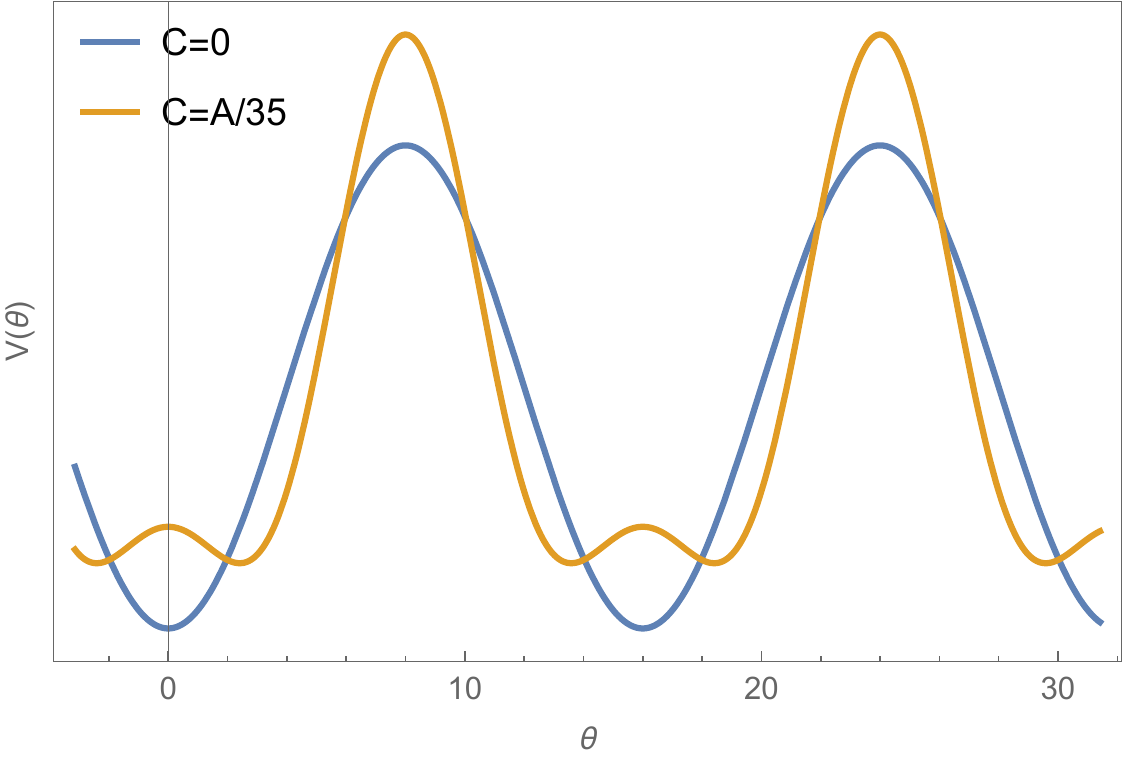}
    \caption{Subleading corrections to axion potential \eqref{eq:subleadAxion} (in arbitrary units) as described in the text for $(A,B,C)=(1,-A/20,A/35)$, $(\alpha,\beta,\gamma)=(2\pi/16, 2\alpha,3\alpha)$. The minima become a maxima when the subleading correction is turned on. }
    \label{fig:cos_mcmc}
\end{figure}

\newpage

\section{Constraints for all the parameters of our models}\label{app:fullmcmc}

In this section we present the plots and tables with the full set of parameters for the three hilltop models we considered in the main text, as well as the Dutta-Scherrer (DS) parameterisation. 

\subsection{Axion hilltop}

\begin{figure}[H]
    \centering
    \includegraphics[width=0.95\linewidth]{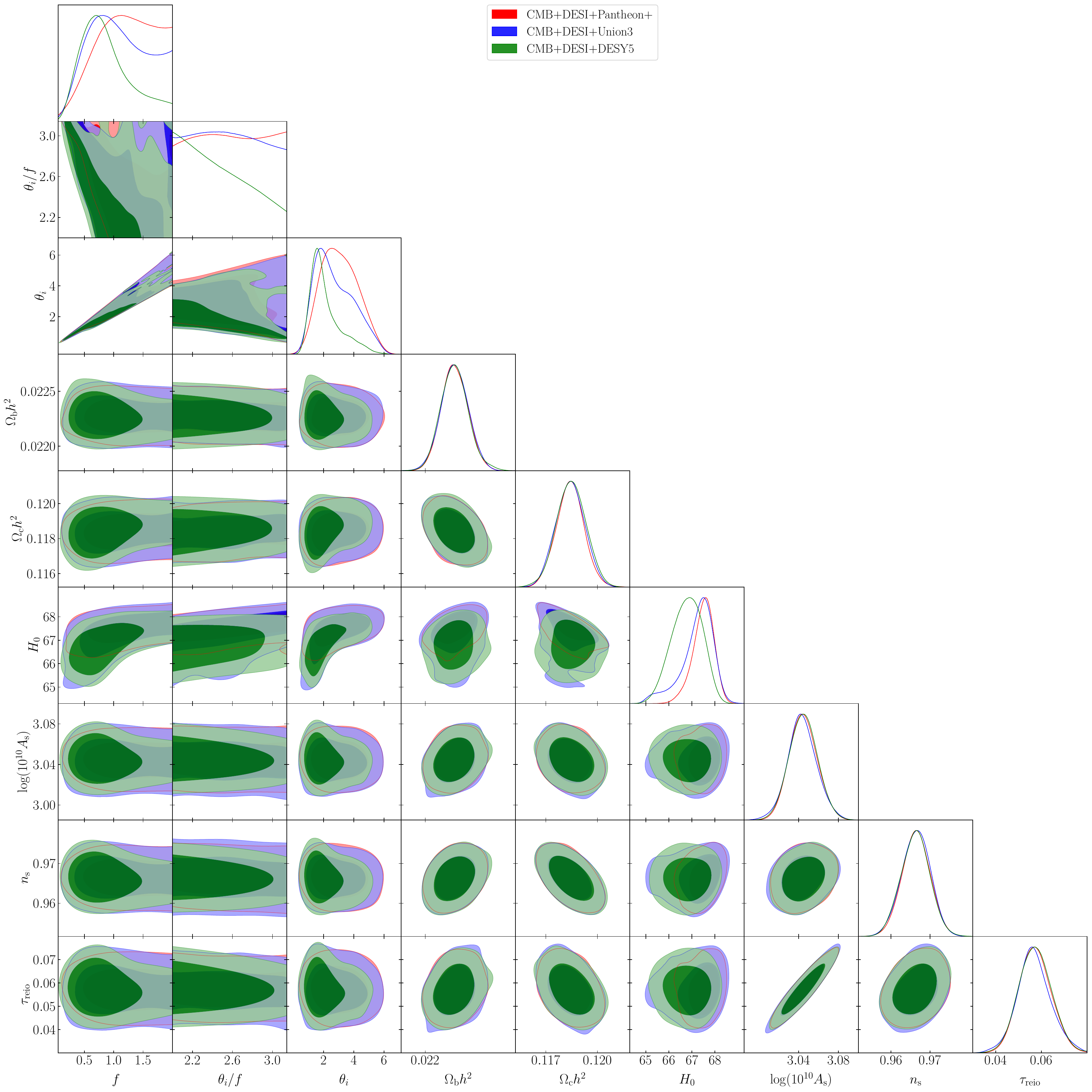}
    \caption{The complete parameter constraints for the Axion model.}
    \label{fig:cos_mcmc}
\end{figure}

\begin{table}[]
    \centering
\begin{tabular} {|c| c| c| c|}

\hline
\rowcolor{gray!30} 
 \bf{Parameter} &  \textbf{+Pantheon+} &  \textbf{+Union3} &  \textbf{+DESY5}\\
\hline
{\boldmath$f              $} & $> 0.946  \ml{1.22}                 $ & $> 0.779   \ml{0.15}                $ & $0.88^{+0.24}_{-0.54} \ml{0.19}      $\\

{\boldmath$\theta_i/f     $} & ---      \ml{2.25}                    & ---     \ml{3.12}                     & $< 2.62    \ml{3.07}                $\\

{\boldmath$\Omega_\mathrm{c} h^2$} & $0.11842\pm 0.00081  \ml{0.11184}      $ & $0.11842\pm 0.00083  \ml{0.118}      $ & $0.11847\pm 0.00086 \ml{0.118}       $\\

{\boldmath$\log(10^{10} A_\mathrm{s})$} & $3.045\pm 0.014  \ml{3.046}          $ & $3.044\pm 0.015 \ml{3.048}           $ & $3.045\pm 0.014 \ml{3.053}           $\\

{\boldmath$n_\mathrm{s}   $} & $0.9664\pm 0.0036  \ml{0.9647}       $ & $0.9664\pm 0.0037  \ml{0.9624}        $ & $0.9663\pm 0.0037   \ml{0.967}       $\\

{\boldmath$H_0            $} & $67.49^{+0.51}_{-0.37}  \ml{67.35}   $ & $67.23^{+0.81}_{-0.40} \ml{66.24}    $ & $66.79^{+0.74}_{-0.62} \ml{65.95}    $\\

{\boldmath$\Omega_\mathrm{b} h^2$} & $0.02227\pm 0.00012  \ml{0.02223}      $ & $0.02227\pm 0.00013  \ml{0.0221}      $ & $0.02227\pm 0.00013 \ml{0.0226}       $\\

{\boldmath$\tau_\mathrm{reio}$} & $0.0573^{+0.0066}_{-0.0074} \ml{0.0592}$ & $0.0568^{+0.0067}_{-0.0076} \ml{0.0586}$ & $0.0573\pm 0.0071 \ml{0.061}         $\\

$\theta_i                  $ & $3.1^{+1.1}_{-1.4}   \ml{2.75}      $ & $2.73^{+0.93}_{-1.6} \ml{0.483}      $ & $2.11^{+0.40}_{-1.2} \ml{0.591}       $\\

\hline
\end{tabular}
    \caption{Axion model: full parameter means and $68\%$ limits for the addition of the different supernovae datasets to the CMB+DESI combination. The values in parentheses denote the best-fit parameters for this model. }
    \label{tab:Axion_table_full}
\end{table}

\newpage

\subsection{Saxion hilltop}

\begin{figure}[H]
    \centering
    \includegraphics[width=0.95\linewidth]{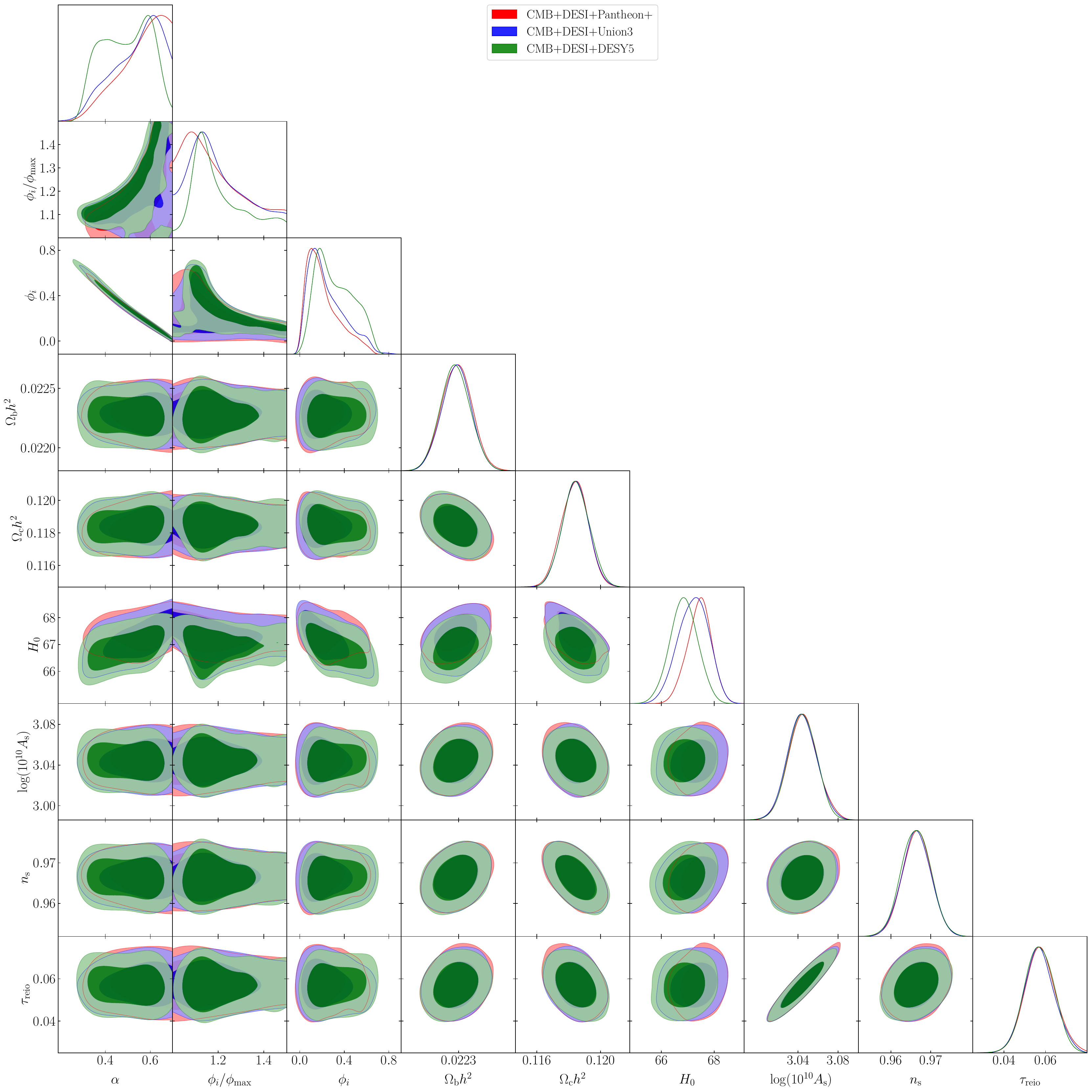}
    \caption{Full parameter constraints for Saxion model.}
    \label{fig:sugra_mcmc}
\end{figure}

\begin{table}[]
    \centering
\begin{tabular} {|c|  c| c| c|}

\hline
\rowcolor{gray!30} 
 \bf{Parameter} &  \textbf{+Pantheon+} &  \textbf{+Union3} &  \textbf{+DESY5}\\
\hline
{\boldmath$\alpha         $} & $>0.521 \ml{0.66}             $ & $0.537^{+0.16}_{-0.045}\ml{0.366}    $ & $0.49\pm 0.11                \ml{0.311}  $\\

{\boldmath$\phi_{i}/\phi_{\mathrm{max}} $} & $< 1.23   \ml{1.43}                  $ & $1.201^{+0.071}_{-0.17} \ml{1.14}   $ & $1.205^{+0.058}_{-0.14} \ml{1.10}   $\\

{\boldmath$\Omega_\mathrm{c} h^2$} & $0.11836\pm 0.00084  \ml{0.1184}      $ & $0.11841\pm 0.00083 \ml{0.1181}       $ & $0.11841\pm 0.00085  \ml{0.1175}      $\\

{\boldmath$\Omega_\mathrm{b} h^2$} & $0.02228\pm 0.00013  \ml{0.02228}      $ & $0.02228\pm 0.00012  \ml{0.02223}      $ & $0.02228\pm 0.00013   \ml{0.02231}    $\\

{\boldmath$\log(10^{10} A_\mathrm{s})$} & $3.045\pm 0.014   \ml{3.040}         $ & $3.044\pm 0.014    \ml{3.048}        $ & $3.044\pm 0.014    \ml{3.045}        $\\

{\boldmath$n_\mathrm{s}   $} & $0.9665\pm 0.0037     \ml{0.9650}     $ & $0.9664\pm 0.0037 \ml{0.9644}         $ & $0.9664\pm 0.0037  \ml{0.9665}        $\\

{\boldmath$H_0            $} & $67.44^{+0.49}_{-0.43} \ml{67.61}    $ & $67.21^{+0.62}_{-0.55} \ml{66.16}    $ & $66.86\pm 0.52       \ml{66.31}         $\\

{\boldmath$\tau_\mathrm{reio}$} & $0.0573\pm 0.0070  \ml{0.0552}        $ & $0.0568^{+0.0067}_{-0.0075}$ \ml{0.0574} & $0.0572^{+0.0066}_{-0.0076} \ml{0.0572} $\\

$\phi_i                    $ & $0.209^{+0.064}_{-0.18} \ml{0.07}     $ & $0.241^{+0.083}_{-0.21}     \ml{0.53}         $ &  $0.32^{+0.13}_{-0.22} \ml{0.64}     $\\
\hline
\end{tabular}
    \caption{Saxion model: full parameter means and $68\%$ limits for the addition of the different supernovae datasets to the CMB+DESI combination.  The values in parentheses denote the best-fit parameters for this model.}
    \label{tab:Sugra_table_full}
\end{table}

\newpage

\subsection{Higgs-like hilltop}

\begin{figure}[H]
    \centering
    \includegraphics[width=0.95\linewidth]{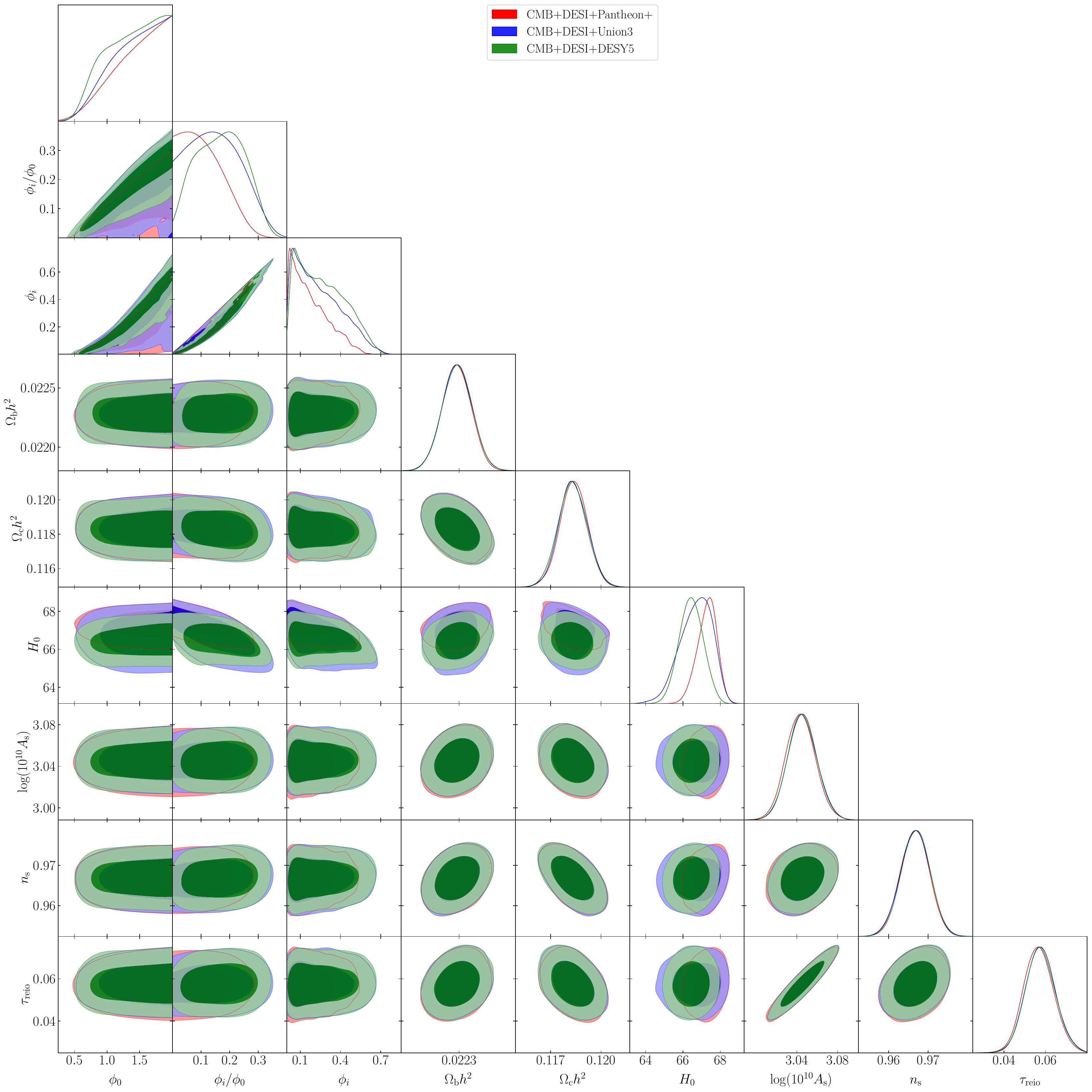}
    \caption{Full parameter constraints for the Higgs-like model.}
    \label{fig:Higgs_mcmc}
\end{figure}

\begin{table}[]
    \centering
\begin{tabular} {|c| c| c| c|}

\hline
\rowcolor{gray!30} 
 \bf{Parameter} &  \textbf{+Pantheon+} &  \textbf{+Union3} &  \textbf{+DESY5}\\
\hline
{\boldmath$\phi_0         $} & $> 1.29    \ml{1.23}                $ & $> 1.24  \ml{0.69}                  $ & $> 1.17  \ml{0.62}                  $\\

{\boldmath$\phi_i/\phi_0  $} & $< 0.142  \ml{0.106}                 $ & $0.151^{+0.073}_{-0.12}  \ml{0.051}  $ & $0.169\pm 0.081  \ml{0.034}          $\\

{\boldmath$\Omega_\mathrm{c} h^2$} & $0.11838\pm 0.00082 \ml{0.1188}       $ & $0.11835\pm 0.00084   \ml{0.1176}     $ & $0.11829\pm 0.00084  \ml{0.1189}      $\\

{\boldmath$\Omega_\mathrm{b} h^2$} & $0.02228\pm 0.00012   \ml{0.02221}     $ & $0.02228\pm 0.00013 \ml{0.00223}       $ & $0.02228\pm 0.00013  \ml{0.02223}      $\\

{\boldmath$\log(10^{10} A_\mathrm{s})$} & $3.044\pm 0.014 \ml{3.041}           $ & $3.046\pm 0.014     \ml{3.044}       $ & $3.045\pm 0.014    \ml{3.033}        $\\

{\boldmath$n_\mathrm{s}   $} & $0.9666\pm 0.0036   \ml{0.9641}      $ & $0.9667\pm 0.0037  \ml{0.9694}        $ & $0.9669\pm 0.0036  \ml{0.9645}         $\\

{\boldmath$H_0            $} & $67.29^{+0.59}_{-0.45} \ml{66.98}    $ & $66.7^{+1.0}_{-0.70}     \ml{66.30}  $ & $66.44\pm 0.64   \ml{66.06}          $\\

{\boldmath$\tau_\mathrm{reio}$} & $0.0567\pm 0.0071  \ml{0.0555}        $ & $0.0576\pm 0.0071       \ml{0.056}   $ & $0.0576\pm 0.0072  \ml{0.0514}        $\\

$\phi_i                  $ & $0.174^{+0.071}_{-0.17} \ml{0.1313}     $ & $0.235^{+0.088}_{-0.23}  \ml{0.0035}  $ & $0.26^{+0.10}_{-0.24}  \ml{0.0021}    $\\
\hline
\end{tabular}
    \caption{Higgs-like model: full parameter means and $68\%$ limits for the addition of the different supernovae datasets to the CMB+DESI combination. The values in parentheses denote the best-fit parameters for this model.}
    \label{tab:Higgs_table_full}
\end{table}

\bigskip
\subsection{Dutta-Scherrer parameterisation}
\begin{figure}[H]
    \centering
    \includegraphics[width=0.95\linewidth]{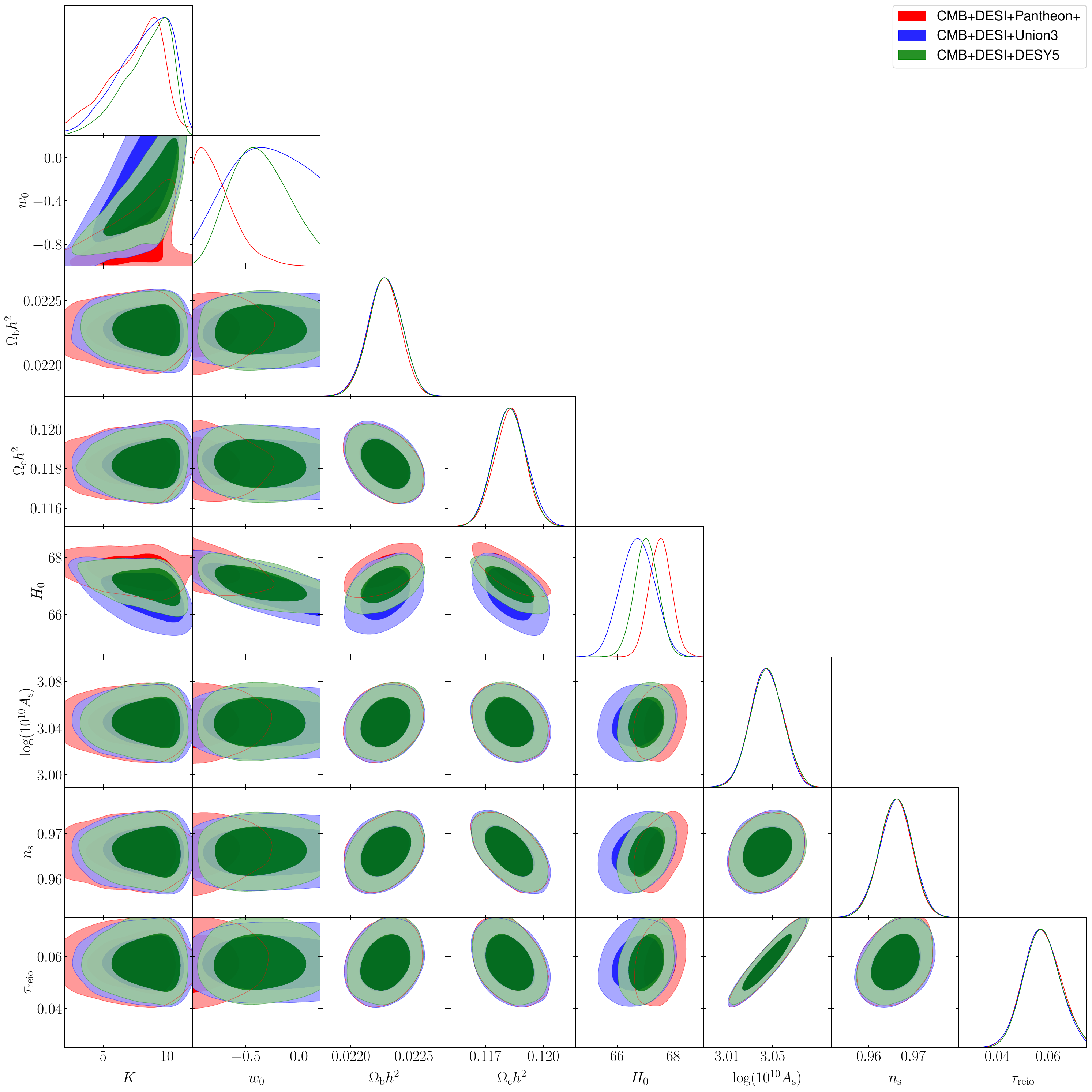}
    \caption{Full parameter plots for the DS parameterisation.}
    \label{fig:DSfull}
\end{figure}

\begin{table}[]
    \centering
\begin{tabular} {|c| c| c| c|}

\hline
\rowcolor{gray!30} 
 \bf{Parameter} &  \textbf{+Pantheon+} &  \textbf{+Union3} &  \textbf{+DESY5}\\
\hline
{\boldmath$K$} & $ 7.6^{+2.5}_{-2.1}              $ \ml{7.49} & $8.2^{+2.7}_{-1.4} \ml{6.16} $ &  $8.4^{+2.4}_{-1.1} \ml{9.52}$\\

{\boldmath$w_0$} &   $<-0.709$    \ml{-0.79}    &   $-0.11^{+0.39}_{-0.61} \ml{-0.60}$      & $-0.35^{+0.24}_{-0.33}$ \ml{-0.21}     \\

{\boldmath$\Omega_\mathrm{c} h^2$} & $0.11830\pm 0.00081         $ \ml{0.11782}& $0.11828 \pm 0.000856   \ml{0.11815}  $ & $0.11825\pm 0.00082 $ \ml{0.11757} \\

{\boldmath$\Omega_\mathrm{b} h^2$} & $0.02226\pm 0.00013$ \ml{0.02228}& $0.02227\pm 0.00013$ \ml{0.02229}& $ 0.02227\pm 0.00013$ \ml{0.02230}\\
{\boldmath$H_0            $} & $ 67.53\pm0.39 $ \ml{67.76} & $66.73\pm 0.59  \ml{65.92}   $ & $67.04\pm 0.42 $ \ml{67.23}\\

{\boldmath$\log(10^{10} A_\mathrm{s})$} & $3.045\pm 0.014            $ \ml{3.046}& $3.044\pm 0.014     \ml{3.045}       $ & $3.045\pm 0.014            $ \ml{3.043}\\

{\boldmath$n_\mathrm{s}   $} & $0.9659\pm 0.0037          $ \ml{0.9687}& $0.9661\pm 0.0037  \ml{0.9676}        $ & $0.9661\pm 0.0036          $ \ml{0.9665}\\

{\boldmath$\tau_\mathrm{reio}$} & $0.0574\pm 0.0071          $ \ml{0.0588}& $0.0573\pm 0.0070    \ml{0.0580}      $ & $0.0576\pm 0.0069          $ \ml{0.0565}\\

\hline
\end{tabular}
    \caption{DS parameterisation: full parameter means and $68\%$ limits for the addition of the different supernovae datasets to the CMB+DESI combination.  The values in parentheses denote the best-fit parameters for this model.}
    \label{tab:DS_table_full}
\end{table}
\begin{figure}[H]
    \centering
    \includegraphics[width=0.7\linewidth]{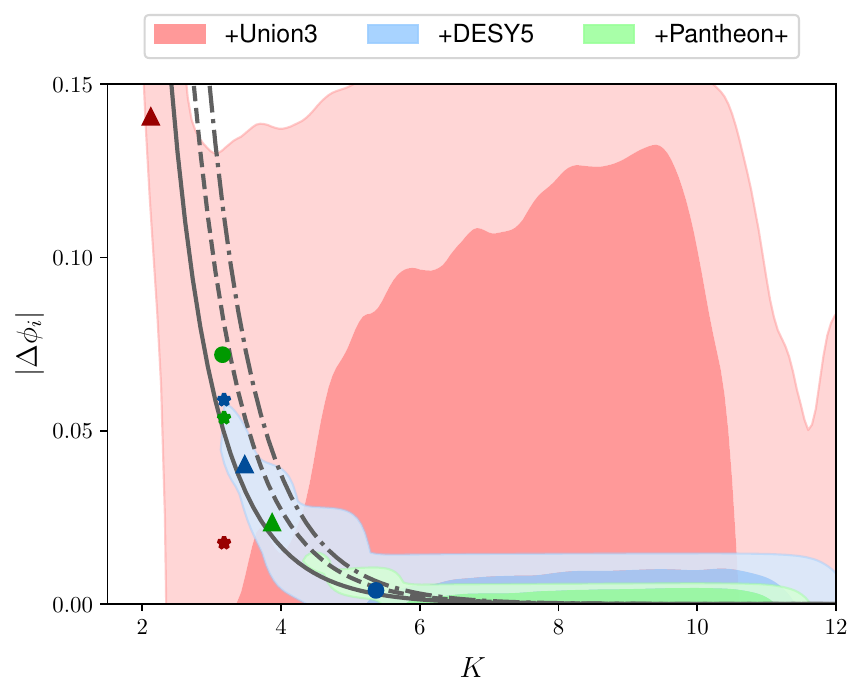}
    \caption{{Analytical result from eq.~\eqref{eq:phii}, results for the hilltop quintessence models and posterior contours from MCMC analysis in the $K$-$\vert \Delta\phi_{i}\vert $ plane. The analytical result is in grey lines, where three different patterns use best-fit values for $\Omega _{\phi,0}$ and $w_0$ from different data combinations (solid: +Pantheon+, dashed: +Union3, dot-dashed: +DESY5). In the same figure, we show the $1\sigma$ and $2\sigma$ bounds from the constraints on the DS parameterisation for all data combinations.
    Values in the $K$-$\vert \Delta\phi_{i}\vert $ plane corresponding to the hilltop quintessence models are indicated by coloured shapes: circle for axion model, star for sugra model and triangle for the Higgs-like model, using 
    best-fit values with Union3/Pantheon+/DESY5 for model parameters
    ($\phi _0$ and $f$) denoted by dark blue/dark red/dark green shapes.  All the best-fit values can be found in Tables \ref{tab:Axion_table_full}-\ref{tab:DS_table_full}. }
    }
    \label{fig:phiK_DS_all}
\end{figure}
\begin{figure}[H]
    \centering
    \includegraphics[width=0.99\linewidth]{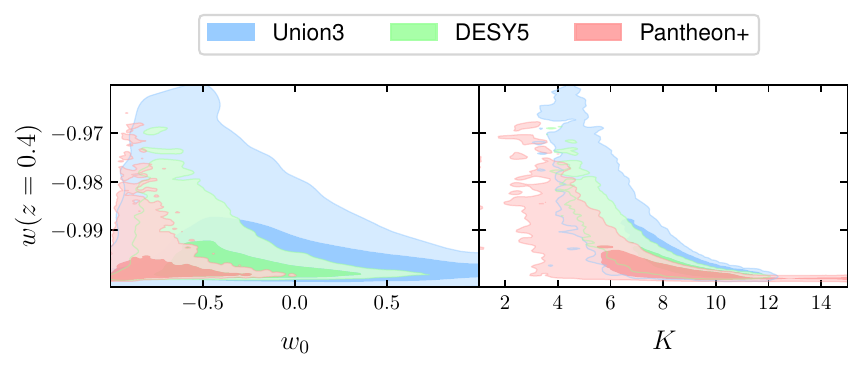}
    \caption{Derived bounds on $w_{\phi} (z=0.4)$ from the MCMC analysis for the DS parameterisation. 
    }
    \label{fig:wzp4_double}
\end{figure}    
\end{appendix}

\newpage

\section{Dutta-Scherrer-Chiba parameterisation, including curvature}\label{app2}

In this appendix we collect the evolution of the equation of state including non-zero curvature, $\Omega_k$ and its comparison with the DS parameterisation, for all the models discussed in the main text. 
In all plots we take $\Omega_k= 0.005$.

\begin{figure}[H]
    \centering
    \includegraphics[width=0.90\linewidth]{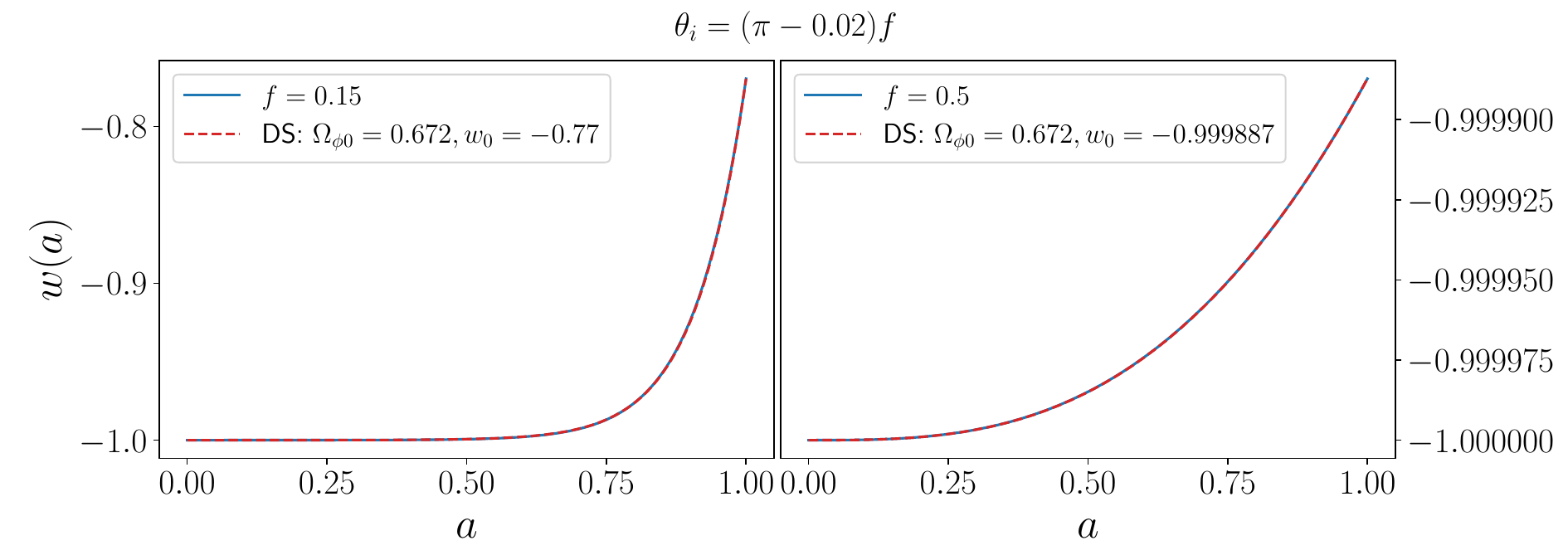}
    \caption{Evolution   of $w_{\rm DE}$ for the Axion model including non-zero curvature and its comparison with the DS parameterisation. }
    \label{fig:cos_curv}
\end{figure}

\begin{figure}[H]
    \centering
    \includegraphics[width=0.65\linewidth]{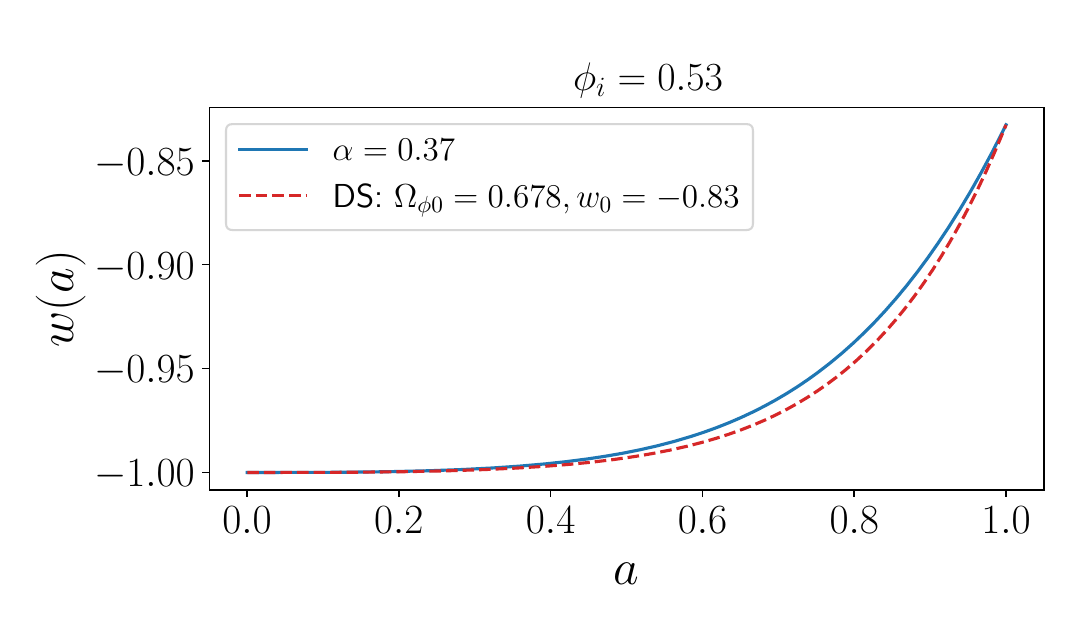}
    \caption{Evolution   of $w_{\rm DE}$ for the Saxion model including non-zero curvature and its comparison with the DS parameterisation. }
    \label{fig:sugra_curv}
\end{figure}

\begin{figure}[H]
    \centering
    \includegraphics[width=0.90\linewidth]{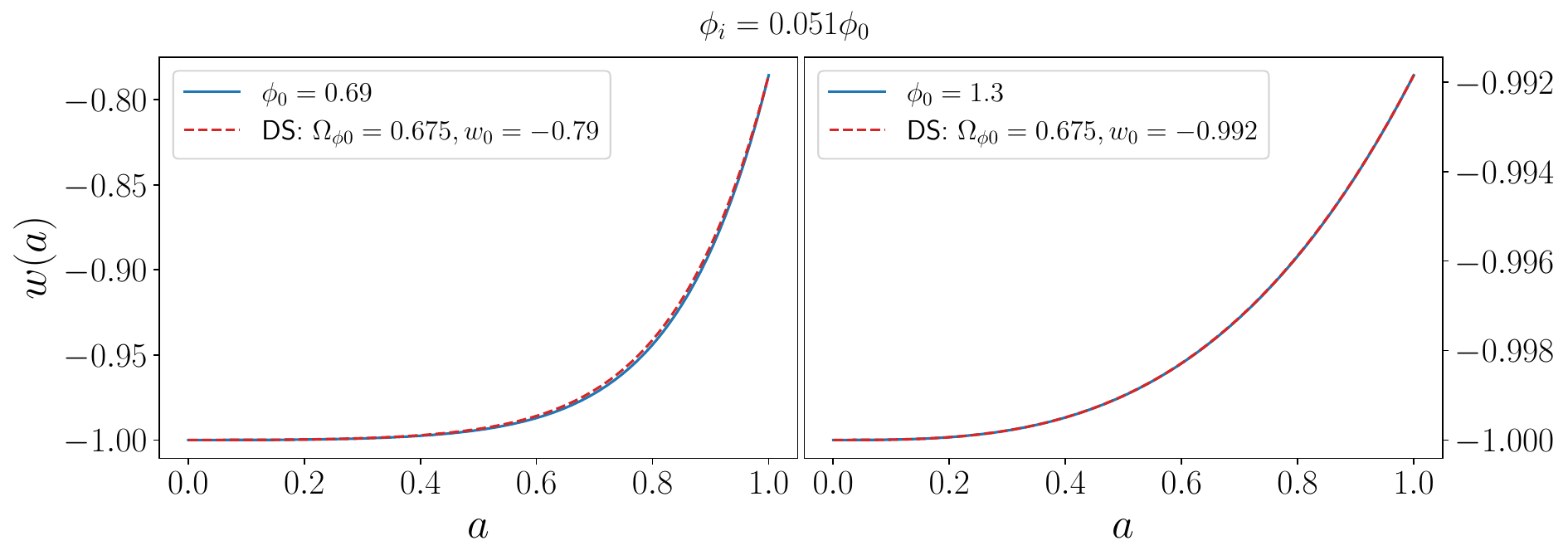}
    \caption{Evolution   of $w_{\rm DE}$ for the Higgs-like model including non-zero curvature and its comparison with the DS parameterisation. }
    \label{fig:FT_curv}
\end{figure}

\begin{figure}[H]
    \centering
    \includegraphics[width=0.9\linewidth]{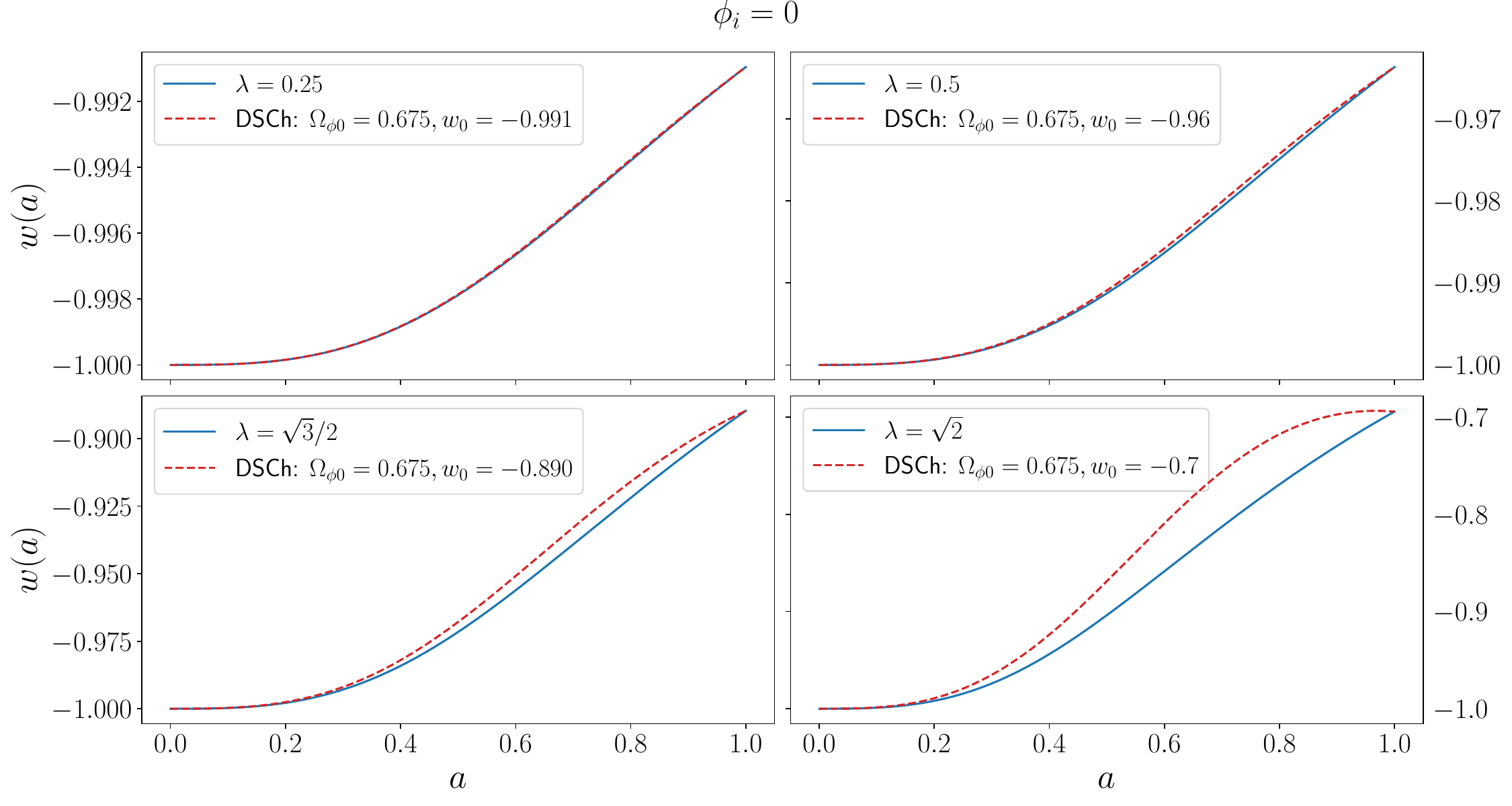}
    \caption{Evolution   of $w_{\rm DE}$ for the Exponential model including non-zero curvature and its comparison with the DSCh parameterisation.}
    \label{fig:exp_curv}
\end{figure}

\newpage

\addcontentsline{toc}{section}{References}
\bibliographystyle{utphys}

\bibliography{refsHill}

\end{document}